\documentclass[letterpaper]{article}

\usepackage[T1]{fontenc}

\usepackage{geometry}
\usepackage{setspace}
\usepackage[disable]{todonotes}
\usepackage[style = chem-acs, articletitle=true]{biblatex}
\usepackage{graphicx}
\usepackage{float}
\usepackage{custom_commands}
\newfloat{scheme}{htbp}{los}
\floatname{scheme}{Scheme}
\floatname{chart}{Chart}
\newfloat{graph}{htbp}{loh}

\usepackage{chemformula} 
\usepackage[version = 4]{mhchem} 
\usepackage{hyperref}
\usepackage{xcolor}
\usepackage{bm}
\usepackage{makecell}
\usepackage{physics}
\usepackage{caption}
\usepackage{authblk}
\author[1]{Dahvyd Wing}
\author[2]{Mihail Bogojeski}
\author[1]{Szabolcs Goger}
\author[2,4,5]{Klaus-Robert M\"uller}
\author[1]{Alexandre Tkatchenko*}
\affil[1]{Department of Physics and Materials Science, University of Luxembourg, L-1511 Luxembourg City, Luxembourg}
\affil[2]{BIFOLD and Machine Learning Group, Technische Universit\"at Berlin, Franklinstr.\ 28/29, 10587 Berlin, Germany}
\affil[4]{Department of Artificial Intelligence, Korea University, Seoul 02841, Korea}
\affil[5]{Max-Planck-Institut f\"ur Informatik, 66123 Saarbr\"ucken, Germany}

\title{Accurate and Transferable Intermolecular Potential Based on Machine-Learned Molecular Electron Density}

\date{*Email: alexandre.tkatchenko@uni.lu}

\begin{document}

\maketitle

\begin{abstract}
Machine-learned force fields (MLFFs) contain many learnable parameters and therefore require large training datasets. This poses a challenge for developing highly accurate, general-purpose MLFFs because generating high-quality ab initio reference data is computationally expensive. Classical empirical potentials offer a potentially inexpensive source of synthetic training data, but existing models often lack the accuracy needed to provide useful reference energies. Here, we introduce the density-based intermolecular potential (DensIP), a physics-based model of intermolecular interactions that uses machine-learned electron densities and only four universal parameters. We train and test DensIP on CCSD(T)/CBS interaction energies from DES15K, a dataset of dimers of small organic molecules. DensIP achieves sub-kcal/mol errors for dimers containing molecules absent from the training set, including molecules in non-equilibrium conformations, demonstrating strong transferability. We further show that DensIP can be applied to molecules as large as drug ligands. Notably, DensIP outperforms state-of-the-art general-purpose MLFFs for long-range interactions, making it a promising approach for generating accurate synthetic training data at scale.
\end{abstract}






\section{Introduction}

Accurate prediction of noncovalent interactions \cite{stone_2013_book} in broad regions of chemical space is essential for many applications such as protein-ligand binding \cite{2010medicinal}, crystal structure prediction \cite{2021_csp_review}, and more. After decades of research, this is still an open challenge for classical force fields \cite{2024sampl9,2025casp16,2019d3r}. Recently, general-purpose machine-learned force fields \cite{mace-off,mace_mp,  mace-omol, mace-polar, so3lr, uma} have been proven to be capable of producing accurate intermolecular interactions with respect to the reference method on which they are trained \cite{tretiakov2025NCI_review}, provided they have enough data \cite{unke2021mlff}. State-of-the-art MLFF models are trained on millions of data points \cite{mace-omol, mace-polar, so3lr, uma}, including forces, at the level of density functional theory (DFT)\cite{parr_and_yang, martin_electronic_structure_book}. However, a general rule of thumb is that atomistic applications require an accuracy of less than 1 kcal/mol, which requires methods that are more accurate than DFT, such as the gold-standard ab-initio method, coupled-cluster with perturbative triples extrapolated to the complete basis set limit, CCSD(T)/CBS \cite{ccsdt}. To date, the largest dataset for CCSD(T)/CBS is 370,000 configurations \cite{des370k}, due to computational expense. Thus, there is not enough high-accuracy data to employ straightforward methodologies of machine-learning (ML) to train general-purpose MLFF models, and given the constraints on computational expense of generating a large CCSD(T)/CBS dataset, alternative solutions must be found.

Many strategies are currently being pursued to this end, which can broadly be divided into methods that try to reduce the amount of data ML models need, such as transfer learning \cite{ani1ccx, transfer_learning} and delta-learning on top of ab-initio methods \cite{delta_ML_water_michaelides_2025, bowman2022delta, bogojeski2020quantum}, and methods that attempt to produce large datasets to be used to train general-purpose MLFFs, such as improving existing ab-initio methods. For the latter strategy, being able to generate a large amount of synthetic data with a high-accuracy empirical force field is particularly attractive, due to empirical force fields' low computational cost. However, developing an accurate and transferable empirical force field is challenging, as noted above. One path forward is to develop empirical force fields that use ML generated molecular properties as components. Such a strategy overcomes the data bottleneck for general MLFFs because empirical force fields use a modest number of parameters, and thus require significantly less data to fit interaction energies, and the additional ML component predicting individual molecular properties also requires less data because learning molecular properties is a combinatorially smaller task than interaction energies and forces between molecules. The challenge is to find a functional form of an empirical force field that is accurate and transferable.

Efforts to construct accurate force fields have used monomer quantities derived from \textit{Ab initio} calculations as components of the force field \cite{qmdff,cole2016, 2016van_vleet_slater_isa, vandenbrande2017MEDFF, van_vleet2018mastiff, gem,fflux, xu2018_fragment_method_review}. In particular, an isotropic, atom-centered representation of the electron density in terms of atom centered charge multipoles and the width of the exponential tail of the density was used \cite{2016van_vleet_slater_isa, vandenbrande2017MEDFF,van_vleet2018mastiff} as determined by the minimal basis iterative stockholder method (MBIS) \cite{mbis}. This has enabled more accurate, physically motivated functional forms to be used for electrostatics, exchange repulsion, and other short-range interactions. Later, it was demonstrated that ML models could be employed to learn these quantum mechanical quantities and eliminate the need for a DFT reference calculation per molecule \cite{2018ipml, 2021cliff}. However, even with more accurate functional forms based on this coarse-grained representation of the density, it has been difficult to achieve sub-kcal/mol accuracy that is transferable across a broad class of molecules \cite{vandenbrande2017MEDFF, 2018ipml} and many atom-type parameters are often employed to compensate for this difficulty \cite{2016van_vleet_slater_isa,  2021cliff}. In order to diagnose the source of the error, symmetry adapted perturbation theory (SAPT) \cite{jeziorski1994sapt_review, stone_2013_book} is often used. SAPT calculates dimer interaction energies as the sum of exchange repulsion, electrostatic, induction, and dispersion interactions. Comparison of force field components with SAPT components revealed that the electrostatic and exchange repulsion components have the highest errors for these models \cite{2021cliff, vandenbrande2017MEDFF, 2016van_vleet_slater_isa}. Thus, we primarily focus our attention on improving these components.

Recently, significant strides have been made in learning electronic structure, in the form of electron density, electronic wavefunctions and Hamiltonians~\cite{rupp2012fast,snyder2012finding,li2014understanding,brockherde2017bypassing,bogojeski2020quantum,bai2022machine,shao2023machine,kirkpatrick2021pushing,huang2023dft,kaniselvan2025helm,khan2025adapting}. Specifically, equivariant models have significantly improved performance~\cite{thomas2018tensor,weiler20183d,unke2021se}, including a series of models capable of predicting the electron density with high precision~\cite{grisafi2018transferable, grisafi2022electronic,rackers2022cracking,jorgensen2022equivariant,bogojeski2023machine,koker2024higher,li2024superres,fu2024recipe,elsborg2026electra}. In some cases, ML predicted densities are even more accurate than direct density fitting \cite{baerends1973_density_fitting,dunlap1979_density_fitting, skylaris2000density_fitting} of the ab-initio density using a conventional density-fitting auxiliary basis set \cite{bogojeski2023machine}. 

Here, we take advantage of an equivariant ML model for the electron density, called DenSNet~\cite{bogojeski2026enhancing}, to develop an empirical intermolecular potential with components that take advantage of having a more accurate representation of the density to achieve better accuracy and transferability. The ML density can be directly used to calculate the electrostatic energy with higher accuracy than what has been achieved by using atomic charge multipoles with a charge penetration model. Additionally, previous work has resulted in the development of an anisotropic valence density overlap (AVDO) exchange repulsion model, which is transferable, uses only 2 universal parameters, and has a factor of two lower error than previous all-electron density overlap models \cite{avdo}. AVDO uses a partial valence density which is defined by removing low-energy molecular orbitals. In this work, we show that the valence density can be machine-learned effectively and can be incorporated into an intermolecular potential. We show that the combination of these two improvements enables the development of a transferable intermolecular potential, which we call DensIP. It contains just 4 universal parameters, instead of tens of atom-type parameters. We fit DensIP on a small dataset of one thousand CCSD(T) interaction energies and show that it achieves excellent transferability across small, closed shell, neutral molecules composed of H, C, N, and O including both relaxed and non-equilibrium conformations of the molecules. DensIP achieves accuracies comparable to state-of-the-art machine-learned general-purpose MLFFs for predicting protein fragment --ligand interactions and outperforms them at predicting long-range interaction energies.

\section{Theory}
We introduce the ML-density intermolecular potential, DensIP, which is inspired by density-based intermolecular potentials \cite{2016van_vleet_slater_isa, vandenbrande2017MEDFF, 2018ipml, 2021cliff}, and, in line with SAPT decomposition analysis, calculates the total interaction energy between a pair of molecules as the sum of electrostatics, exchange repulsion, induction, and dispersion components:
\[ E_{int} = E_{elst} + E_{exch} + E_{ind} + E_{disp}.\]
In the following subsections we will discuss how each of these components is computed.

\subsection{Electrostatics}
The electrostatic energy is computed straightforwardly as:
\[ E_{elst} = \int \frac{\rho_A(r) \rho_B(r')}{\abs{r-r'}}\dd{r} \dd{r'}
- \sum_b \int \frac{\rho_A(r) Z_b}{\abs{r-R_b}}\dd{r}
- \sum_a \int \frac{Z_a \rho_B(r')}{\abs{R_a-r'}} \dd{r'} 
+ \sum_{a,b} \frac{Z_a Z_b}{R_a - R_b},\]
where $\rho_A$ and $\rho_B$ are all-electron densities of molecule A and molecule B. Unlike in charge multipole models, using the all-electron density to compute the electrostatic energy means that the charge penetration energy is accounted for exactly. All electron densities are predicted by a DenSNet model, which we call DenSNet-AE, and represented on a Gaussian basis set (more details are provided in a subsequent section on DenSNet). Electrostatic energies are computed from these densities analytically via \texttt{PySCF} \cite{2018pyscf, 2020pyscf, 2015libcint}.

\subsection{Exchange-Repulsion}
The Pauli exchange repulsion energy can be empirically modeled by the electron density overlap model \cite{kita1976, kim_1981_ovlp}, however, for several decades such models needed to be fit per system \cite{ kim_1981_ovlp, ihm_1990_charge, nobeli1998_different_K, piquemal_darden_2006, elking2010gmm, bygrave_manby2012} or employed many atom type parameters \cite{mitchell_price2000atomtypes, 2016van_vleet_slater_isa, 2021cliff}. Recently, we have shown that the anisotropic, valence density overlap model (AVDO) is a transferable model for exchange repulsion using only two global parameters \cite{avdo}:
\begin{equation}
\label{eq:AVDO}
E_{exch} = K_{exch} \bigg(\int \rho_{A,val} \rho_{B,val}\dd{\bm{r}}\bigg)^{\alpha_{exch}},
\end{equation}
where $K_{exch}$ and $\alpha_{exch}$ are fitted parameters. The model uses valence densities, $\rho_{val}$ which are defined as:
\[\rho_{val} = \sum_{i=M+1}^{N} \abs{\psi_i}^2,\]
where $N$ is the total number of electrons in the monomer, $\psi$ are molecular orbitals, and $M$ is the number of orbitals that are not included in the density. We compute $M$ by $M =  2N_C +  4N_N + 4N_O$, where $N_C$ is the number carbon atoms in the molecule, $N_N$ is the number of nitrogen atoms, and $N_O$ is the number of oxygen atoms. This heuristic method removes low-energy molecular orbitals associated with the 1s and 2s orbitals of second row p-block elements, which was shown to significantly improve the transferability of the model \cite{avdo}. We train a DenSNet model to predict valence densities, which we refer to as DenSNet-val, and compute the overlaps from these valence densities analytically using \texttt{PySCF}.

\subsection{Induction}
Phenomenologically, SAPT induction is a combination of charge-transfer, a short-range effect, and polarization, a long-range effect \cite{stone_2013_book}. However, they appear together as the result of second-order perturbation in the SAPT expression, such that it is not easy to distinguish between them \cite{stone_2009_CT,stone_2013_book}. Several methods have been suggested to separate these two phenomena quantitatively \cite{stone_2009_CT, misquitta2013_regularized_sapt_CT, rezac2015_charge_transfer_cdft, 2016_almo_CT}, however there is no consensus yet as to what is the best way to do this \cite{mao2018_compare_CT}. Even with this quantitative ambiguity however, it is clear that the charge-transfer energy correlates strongly with the exchange repulsion and charge-penetration energies, all of which are short-range phenomena \cite{stone_2009_CT}. Thus, density overlap models have also been used as empirical models for charge-transfer \cite{2016van_vleet_slater_isa, vandenbrande2017MEDFF, 2021cliff}. Initial tests showed that the AVDO method was more accurate than an all-electron model (see the SI \cite{SI}). Thus, similar to other density-based models \cite{2016van_vleet_slater_isa, 2021cliff}, we model the induction component as:
\begin{equation}
\label{eq:ind}
E_{ind} = K_{CT} \bigg(\int \rho_{A,val} \rho_{B,val}\dd{\bm{r}}\bigg)^{\alpha_{CT}} + E_{pol},
\end{equation}
where $K_{CT}$ and $\alpha_{CT}$ are fitted parameters, and $E_{pol}$ is the polarization energy. The valence density used here is the same as the one used in the exchange repulsion model, namely the density predicted by DenSNet-val.

The polarization energy is calculated by a classical, distributed polarization model \cite{thole1981, stone_2013_book} using Gaussian induced dipoles \cite{elking_2007_pol}:
\begin{equation}
\label{eq:E_pol_gauss}
  E_{pol} = \min_{\mu}\left[-\sum_a \vec{E}_{field,a} \cdot \vec{\mu}_a -\frac{1}{2} \sum_{a\ne b}\mu_{a,i} T_{ij}(R_{ab},\beta_{ab}) \mu_{b,j} + \frac{1}{2}\sum_a \tilde{\alpha_{a}}^{-1} \abs{\mu_{a}}^2\right].
\end{equation}
Here, Einstein notation denotes summation over Cartesian coordinates $i$ and $j$, indices $a$ and $b$ both iterate over all atom sites, $R_{ab}$ is the distance between atom sites $a$ and $b$, and $\mu$ is an induced dipole. The electric field at atom site $a$, $E_{field, a}$, is generated by other molecules.  $\tilde{\alpha}$ is the Tkatchenko-Scheffler (TS) rescaled isotropic polarizabilities \cite{TS_rescale_2009} and $\beta_{ab} = (\tilde{\alpha_{a}} \tilde{\alpha_{b}})^{-1/6}$. The Gaussian dipole damping function for the dipole-dipole interaction $T_{ij}$ between two atomic sites is \cite{elking_2007_pol}:

\begin{equation}
\label{eq:T2_gauss}
 T_{ij}(R,\beta) = \frac{3R_i R_j}{R^5}\bigg(\erf(\beta r) - \frac{2}{\sqrt{\pi}} \big( \beta R + \frac{2}{3}(\beta R)^3\big)e^{-\beta^2 r^2}\bigg) -\frac{\delta_{ij}}{R^3} \bigg(\erf(\beta R) - \frac{2\beta R}{\sqrt{\pi}} e^{-\beta^2 R^2}\bigg).
 \end{equation}
To quantify the errors of the model, we compare molecular polarizabilities calculated by density functional perturbation theory with the PBE functional and the polarization model using isotropic atomic polarizabilities computed at the same level of theory. The result is that the model has 15\% errors in predicting molecular polarizabilities. For the rest of this study, we use CCSD(T) atomic polarizabilities \cite{schwerdtfeger2019_atomic_polarizability, hait2018_atomic_polarizability} and electric fields calculated using the densities from DenSNet-AE. TS rescaling uses Hirshfeld volumes ratios \cite{TS_rescale_2009}
, which we calculate by Hirshfeld partitioning \cite{hirshfeld1977} the densities produced from DenSNet-AE. Polarization energies are computed via an in-house code.

\subsection{Dispersion}
The many-body dispersion (MBD) method \cite{mbd} using range-separated self-consistent screening (rs-SCS) \cite{ambrosetti2014_rsscs} has been shown to be an effective model for dispersion \cite{nickerson2023_dispersion_benchmark} which, as its name suggests, is able to capture not only pairwise dispersion between atoms, but also many-body dispersion effects. In addition to requiring atomic positions and element types, MBD requires Hirshfeld volume ratios to rescale C6 coefficients and atomic polarizabilities. As in the polarization model, these are calculated from the densities produced by DenSNet-AE. DensIP MBD energies are calculated using libMBD \cite{libmbd}.

\section{Methodology}
\subsection{DenSNet}
\label{sec:DenSNet}
DenSNet is a graph neural network that uses spherical harmonic representations of each atom's environment that preserve the orientation of the molecule and ensures that all transformations in the network are rotationally equivariant. The model then uses these equivariant atomic representations to predict the coefficients of an atom-centered basis representation of the electron density (see ref.~\cite{bogojeski2026enhancing} for a detailed description of DenSNet's architecture).

The target electron density is represented as a sum of atom-centered Gaussian-type orbital (GTO) basis functions. Given a set of nuclear positions $\Rr = \{\Rr_i | i \in 1\dots M\}$ and charges $\mathrm{Z} = \{Z_i | i \in 1\dots M\}$, the model uses the atom-centered spherical harmonic tensors to predict angular (spherical-harmonic) basis coefficients $\bgamma$, but in contrast to standard density fitting approaches, DenSNet also predicts the Gaussian widths $\balpha$ and radial scale coefficients $\bbeta$. These parameters are collectively denoted as $\bomega \equiv \{\balpha,\bbeta,\bgamma\}$, and they can be used to expand the density $\rho$ on an arbitrary grid as:
\begin{equation}\label{eq:ml_basis_def}
  \rho_{\bomega}(\rv) = \sum_{i=1}^{M}\rho_i(\rv_i),\qquad \rv_i = \rv - \Rr_i,
\end{equation}
where each atom's contribution is an expansion in real spherical harmonics and Gaussian radial functions:
\begin{equation}\label{eq:atom_density}
  \rho_i(\rv_i) = \sum_{l=0}^{L}\sum_{j=1}^{J_{il}} \bgamma_{ij}^{(l)}\cdot Y_l(\uv_i)\;\beta_{ilj}\,A_{ilj}\,e^{-\alpha_{ilj}\,\lVert\rv_i\rVert^2},
\end{equation}
with $\uv_i = \rv_i/\lVert\rv_i\rVert$. Here $Y_l(\uv_i)$ denotes the vector of all $(2l+1)$ real spherical harmonics of degree $l$, the product $\bgamma^{(l)}_{ij}\cdot Y_{l}(\uv_i)$ is a dot product over the $m$ indices and the factor $A_{ilj}$ normalizes each primitive Gaussian.

The fact that DenSNet predicts the Gaussian widths and radial scale coefficients along with the spherical coefficients for each atom allows the radial part of the basis to vary based on the local environment of the atom. This leads to higher basis flexibility, particularly when modeling non-equilibrium geometries.

To reduce the burden of learning the high magnitude, but constant core electron densities, DenSNet can also predict the density as $\rho = \rho_{\mathrm{SAD}} + \rho_{\Delta\mathrm{ML}}$, where $\rho_{\mathrm{SAD}}$ is the superposition of free-atom densities and $\rho_{\Delta\mathrm{ML}}$ is the machine-learned delta correction. This delta-learning strategy focuses the network's capacity on the smoother residual bonding density rather than the largely invariant core region, leading to faster convergence and improved accuracy. Since the core region is important for electrostatic calculations, the all-electron model, DenSNet-AE, uses this delta-learning technique. For the valence density model, DenSNet-val, the core has already been largely removed, so the model directly predicts the density rather than predicting a delta correction.

DenSNet-AE predicts $\rho_{\Delta ML}$ on a basis set where the number of gaussian atomic orbitals per element and angular momentum is set to match the aug-cc-pVQZ-jkfit auxiliary basis \cite{weigend2002_jkfit}. DenSNet-val predicts $\rho$ on a basis set where the number of gaussian atomic orbitals has been custom set to better represent the tails of the valence density (see the SI for further details \cite{SI}). Both models contain 3M trained parameters. We enforce charge conservation on the predicted densities, which is crucial for the accurate computation of interaction energies. Hirshfeld volumes for the induction and dispersion components are calculated by evaluating the density in real space.

\subsubsection{DenSNet training procedure}
Both DenSNet models were trained on QM7-X 250~\cite{hoja2021qm7}, a curated subset of the QM7-X dataset covering 250 organic molecules composed of H, C, N, O, S, and Cl with up to 7 heavy atoms in 100 different conformations computed at the PBE/aug-cc-pVDZ level of theory using \texttt{PySCF}~\cite{frank2022so3krates}. The training set consists of 20,000 geometries, with another 2,500 held out for validation. The training dataset is further limited to only molecules containing H, C, N, and O for training DenSNet-val, as a definition for the valence density of S and Cl has not yet been determined.

Both models are trained by minimizing the absolute fractional error (AFE) between the predicted and reference density, evaluated on angular-radial integration grids:
\begin{equation}\label{eq:perc_dens_diff}
  \mathcal{L}_{\mathrm{dens}}(\{\rho_{\bomega,b}\}_{b=1}^{N_{\mathrm{batch}}},\{\rho_b\}_{b=1}^{N_{\mathrm{batch}}}) = \frac{1}{N_{\mathrm{batch}}}\sum_{b=1}^{N_{\mathrm{batch}}}{\int} \frac{\left|\rho_{\bomega,b}(\rv) - \rho_b(\rv)\right|}{\int\rho_b(\rv)d\rv}d\rv.
\end{equation}
The AFE loss is computed on a random subsample of grid points per training step to reduce computational cost, using 10,000 grid subsamples per molecule.
Parameters were optimized with Adam~\cite{kingma2014adam} using an initial learning rate of $10^{-3}$.
The validation loss was evaluated every 10,000 training steps, and the learning rate was decayed by a factor of 0.5 if the validation loss did not decrease for five consecutive evaluations.
Training was stopped once the learning rate fell below $10^{-5}$, for a maximum of 300,000 steps. The final validation loss of the DenSNet-AE model is 0.0015 and for DenSNet-val it is 0.007.

\subsection{DensIP Parameterization and Testing}
\subsubsection{Datasets}
For parameterizing and testing the transferability of DensIP we use the DES15K dataset \cite{des370k}, a dataset of interaction energies of dimers of small organic molecules with up to 8 heavy atoms computed with coupled-cluster singles and doubles with perturbative triples at the complete basis set limit, CCSD(T)/CBS. We use neutral dimers consisting of 78 unique molecules composed only of H, C, N, and O. The dataset has two subgroups. The first group consists of individually relaxed molecules in optimized dimer configurations sampled at four points along a dissociation curve: an extremely close-range, highly repulsive configuration at $\sim$3 $E_{eq}$ above the minimum energy, $E_{eq}$; a short-range configuration at the zero-crossing point; the minima; and a distant configuration with energy $0.5 E_{eq}$ above the minima. We refer to this subgroup as the optimized dimer set and it contains a total of 4,063 configurations of 1,016 unique dimers. The second group is composed of out-of-equilibrium geometries taken from nearest neighbor dimers from condensed phase MD simulations at 298 K consisting of 74 molecules solvated in water and 56 molecules in a neat liquid (1,580 configurations in total).

To analyze the components of DensIP, we also compute SAPT(DFT)  \cite{williams2001sapt_dft, misquitta2002sapt_dft} and PBE+MBD \cite{ambrosetti2014_rsscs} energies on the optimized dimer dataset using \texttt{Q-Chem} \cite{2021qchem}. The SAPT(DFT) calculations use the PBE functional and an aug-cc-pVDZ basis set so that the electrostatic energy matches the level of theory used to train the DenSNet-AE model. PBE+MBD calculations use an aug-cc-pVTZ basis set. Convergence analysis for these basis sets was previously reported in \cite{avdo}.

We further test the transferability to larger, application-oriented systems by using PLF547 \cite{plf547}, a dataset of dimers of drug ligands and protein fragments similar in size to DES15K. The only neutral dimers containing H, C, N, and O are used, such that the dataset contains 127 dimers of four drug ligands with 30 - 38 heavy atoms. The interaction energies of this dataset are also computed using CCSD(T)/CBS.

\subsubsection{Parameterization} 
We split the optimized dimer dataset into a training dataset composed of dimers of 40 molecules (1,016 calculations) and a test set composed of dimers of a different set of 38 molecules (1,144 calculations). We exclude optimized dimers which are composed of one molecule from the training set and one molecule from the test set. The train-test split employed 3-level stratification where molecules were sorted based on the average CCSD(T) energy of dimers containing that molecule. This ensured that the train and test sets contained similar distributions (see the SI \cite{SI}). The DES15K dataset has five molecules in common with the QM7X-250 dataset used to train the DenSNet models, thus we place dimers containing those molecules in the training dataset, so that the test set is a test for both DenSNet and DensIP. The MD dimers subset is also reserved for testing to demonstrate transferability to different conformations, and it contains molecules from both the training and test sets.

First, we separately fit DensIP's exchange repulsion and induction components to first-order SAPT(DFT) exchange repulsion energies, $E_{exch, SAPT}$, and a reference induction energy given by:
\[ E_{ind, ref} = E_{CCSD(T)} - E_{elst, SAPT} - E_{exch, SAPT} - E_{MBD}.\]
The parameterization minimizes the mean absolute error (MAE) in order to prevent the model from being strongly biased towards the highly repulsive region, which has the largest errors. The exchange repulsion component and induction component are highly correlated and we fit the two components separately to provide a starting point for a global fit. For the global fit, we minimize the following loss function:
\begin{equation}
\label{eq:loss}
 L =  \textrm{MAE}(E_{int}) + w \textrm{MAE}(E_{exch}),
\end{equation}
where $\sigma_{E_{int, ref}}$ is the standard deviation of the CCSD(T)/CBS interaction energies and $w$ is a hyperparameter that constrains the minimization so that the exchange repulsion energy is physically meaningful with respect to SAPT values. The Nelder-Mead algorithm is used to minimize the loss for each of the parameterization steps as implemented in \texttt{SciPy} \cite{2020SciPy}. Hyperparameter optimization of $w$ is performed using 5x5 crossfold validation and is reported in the SI \cite {SI}. While the accuracy of the model changes negligibly with $w$ due to the fact that the exchange repulsion and charge-transfer models are highly correlated, larger $w$ results in significantly better agreement between DensIP's exchange repulsion component and the SAPT(DFT) exchange repulsion energy, and we set $w=0.3$ because there is a slight minimum in the MAE of the exchange-repulsion energy at this value and this also results in reduced variance in the other fitted parameters (see the SI for details \cite{SI}). We also provide a learning curve that demonstrates that a training dataset of 32 molecules approximately saturates the DensIP model in the SI. With these checks in place, we parameterize the model on the full training dataset, resulting in the parameters in Table \ref{tbl:parameters} (we additionally note that for eqs. \ref{eq:AVDO} and \ref{eq:ind} the density is calculated in Bohr$^{-3}$).

\begin{table}
  \caption{DensIP parameters}
  \centering
  \label{tbl:parameters}
  \begin{tabular}{cccc}
  \hline
    $K_{exch}$ (Ha) & $\alpha_{exch}$ & $K_{CT}$ (Ha) & $\alpha_{CT}$\\
    \hline
 25.11 & 0.931 & -1.97 & 0.681\\
    \hline
  \end{tabular}
  \end{table}

\section{Results and Discussion}

\begin{figure}[htbp]
    \centering
        \includegraphics[width=0.9\textwidth]{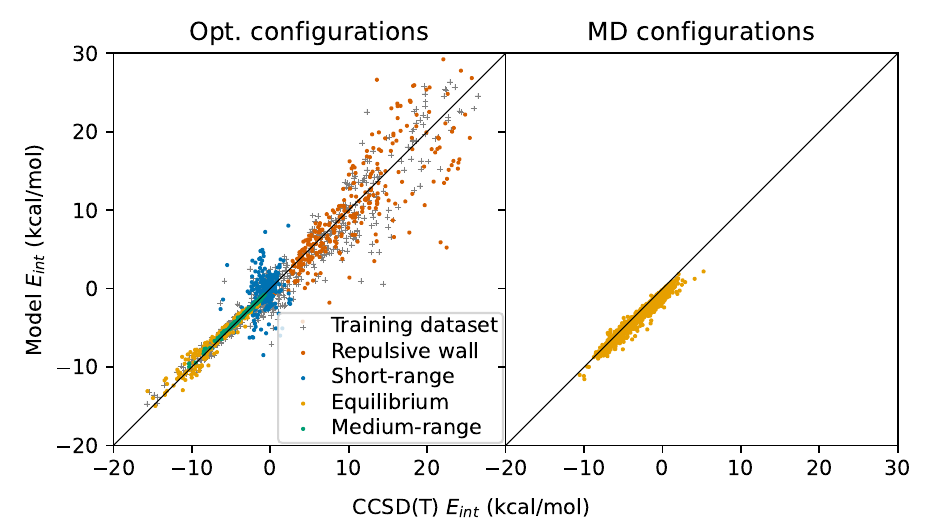}
        \hfill
    \caption{Evaluation of DensIP on DES15K. The training dataset is marked by black crosses and the test datasets are marked by colored dots. In the case of the optimized configurations, we divide the test set into categories according to intermolecular distance. We note that the optimized configuration training and test datasets contain entirely different molecules, demonstrating DensIP's transferability.}
    \label{fig:DES15K}
\end{figure}

We report the accuracy of DensIP on the optimized dimer training and test datasets categorized by dimer intermolecular distance, see Table~\ref{tbl:DES15K_PLF547} and Fig.~\ref{fig:DES15K}. We see that DensIP has only a slight performance drop on the test set, confirming that DensIP is indeed transferrable to small molecules outside of its training set. The error increases as the molecules approach each other, mainly due to errors in the exchange repulsion and induction models, as will be discussed later in Table~\ref{tbl:component_errors}. When evaluated on the MD dimer set, we see that the error is similar to the equilibrium dimer configurations. This confirms that the model is also transferable to non-equilibrium conformations, even when trained only on relaxed conformations. For comparison, we also report errors of leading general-purpose MLFFs on the full DES15K dataset, but note that the SPICE \cite{spice} and OMOL25 datasets \cite{omol25} used to train them include molecules from the DES15K dataset such that it is not clear that DES15K is an appropriate test for these models. We note that this means that some of the general-purpose MLFFs evaluated here were trained on over 100 million DFT calculations compared to DenSNet being trained on 20,000 DFT calculations and DensIP being trained on 1,016 CCSD(T)/CBS calculations.

\begin{table}
  \caption{Interaction Energy RMSE with respect to CCSD(T)/CBS on the DES15K and PLF547 datasets (kcal/mol)}
  \label{tbl:DES15K_PLF547}
  \begin{tabular}{ccccccc}
    \hline
     &   Repulsive wall & Short-range & Equilibrium & Medium-range & MD & PLF-547\\
    \hline
DensIP (train)      & 3.1 & 1.7 & 0.6 & 0.1   & -  & -  \\
DensIP (test)   &   3.8 & 2.2 & 0.7 & 0.2  & 0.7 & 0.7\\
SO3LR v1 \cite{so3lr}         & 2.5 & 1.8 & 1.0 & 0.5 & 0.6 & 0.9 \\
SO3LR v2 \cite{so3lrv2} & 1.6 & 0.9 & 0.6 & 0.3 & 0.5 & 0.4 \\
MACE-OFF24-m  \cite{mace-off}   & 1.0 & 0.6 & 0.3 & 0.2 & 0.2 & 0.4 \\
MACE-POLAR-1m \cite{mace-polar}  &  0.6 & 0.4 & 0.2 & 0.1 & 0.1 & 0.6 \\
MACE-OMOL \cite{mace-omol}   & 0.6 & 0.4 & 0.2 & 0.1 & 0.2 & 0.2 \\
    \hline
  \end{tabular}
  \end{table}

\begin{figure}[htbp]
    \centering
        \includegraphics[width=0.9\textwidth]{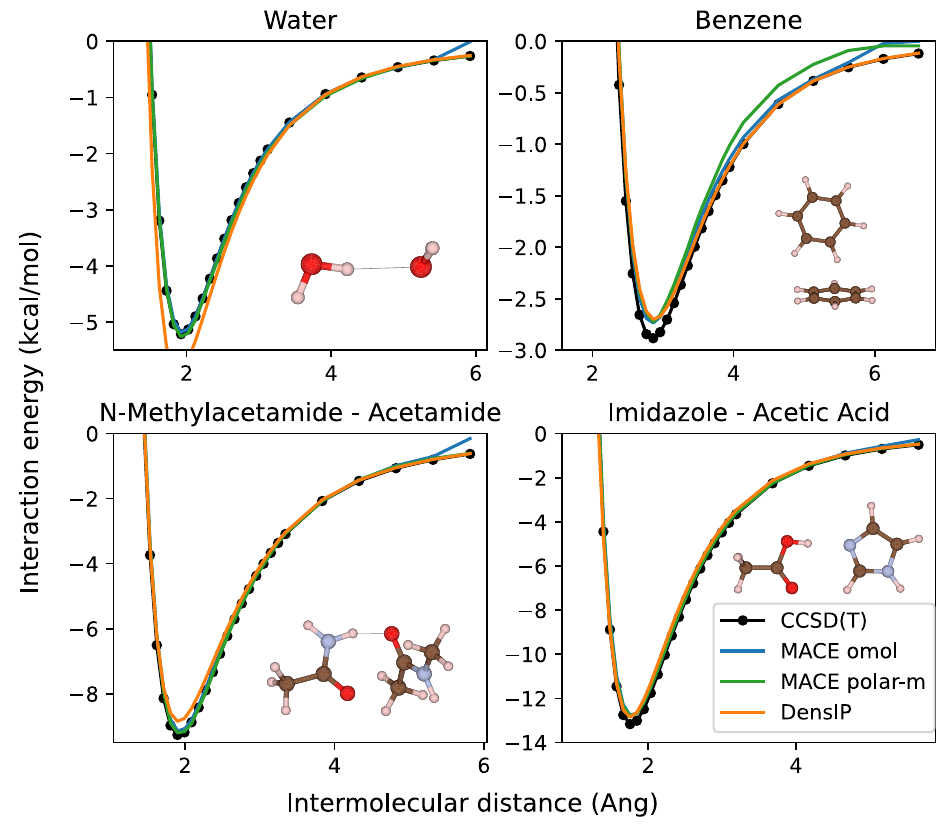}
        \hfill
    \caption{Dissociation curves for four representative dimers from DES15K}
    \label{fig:4_curves}
\end{figure}

\begin{figure}[htbp]
    \centering
        \includegraphics[width=0.75\textwidth]{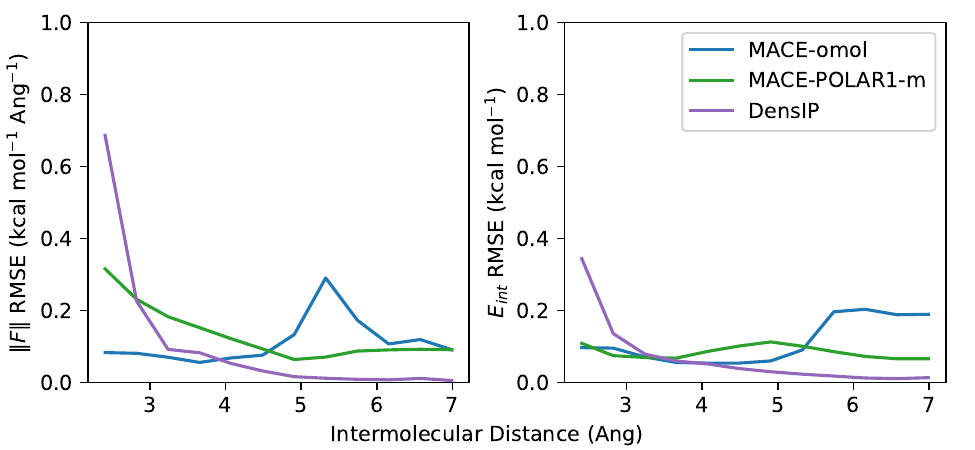}
        \hfill
    \caption{Left) RMSE errors in the magnitude of the net force along the dissociation curves in DES370K \cite{des370k} binned by the nearest atom intermolecular distance (only dimers present in DES15K are used). The dimer configurations shown are farther apart than the medium range configuration in DES15K. Right) Error in the interaction energy for the same dimer configurations, also binned by intermolecular distance. }
    \label{fig:long_range}
\end{figure}

Next, we show four representative dissociation curves for DensIP from the DES15K dataset and compare it to general-purpose MLFFs. The water, N-methylacetamide - acetamide, and imidazole - acetic acid dimers contain hydrogen bonds; the latter two dimers are model protein backbone-backbone interactions and side chain - side chain interactions respectively, see Fig. \ref{fig:4_curves}. The benzene dimer is a prototypical dispersion bound dimer. DensIP predicts smooth dissociation curves that, though not as accurate as MLFFs for near equilibrium configurations, become increasingly accurate in the long-range compared to state-of-the-art MLFFs. MACE-OMOL due to its long-range cutoff function spuriously goes to zero around 6 angstroms. MACE-polar is an improvement in this respect, however it is not uniformly better, as demonstrated by the benzene homodimer. To further demonstrate DensIP's uniformly accurate long-range behavior quantitatively, we calculate the RMSE of the magnitude of the net intermolecular force along the dissociation curves for the entire optimized dimer dataset, binning dimers according to the shortest intermolecular distance between the monomers. The results are shown in Fig. \ref{fig:long_range}. We also plot the RMSE of the interaction energy for each distance bin. The long-range cut off of MACE-OMOL leads to a noticeable peak in the error between five to six Angstroms. The error for MACE-POLAR-1m is better, but nevertheless it does not disappear at long-ranges. On the other hand DensIP's error does go to zero, such that it is negligible by 5 angstroms. While a 0.1 kcal/mol error between two dimers in vacuum is not large, these long-range errors will accumulate in the condensed phase, and given the exponential behavior of the Boltzmann distribution, will likely noticeably bias the dynamics of heterogeneous systems, like protein-ligand pockets. Additionally, large force errors can skew dynamic trajectories and may cause potential energy surfaces to be rough, which is known to affect diffusion coefficients \cite{zwanzig1988diffusion}.

\begin{figure}[htbp]
    \centering
        \includegraphics[width=0.75\textwidth]{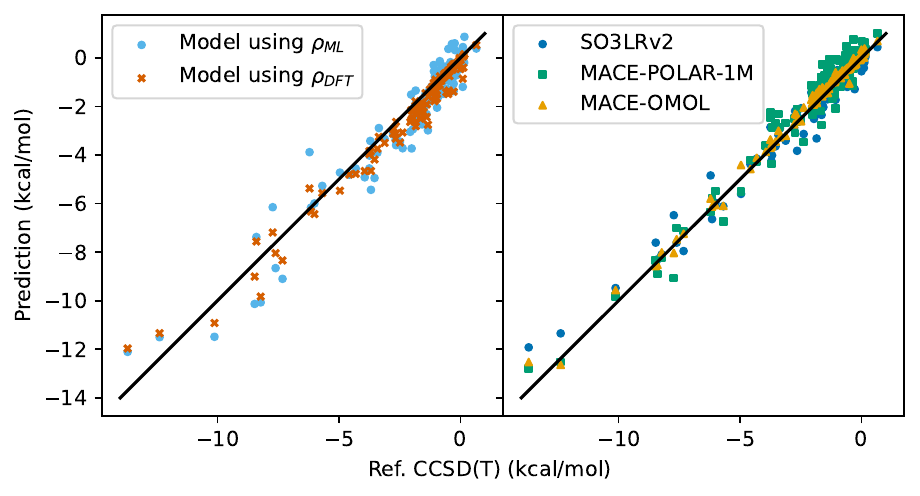}
        \hfill
    \caption{Testing the interaction energy of various models on the PLF547 dataset of ligand-protein fragments.}
    \label{fig:PLF547}
\end{figure}

Next, we investigate DensIP transferability to larger molecules and its applicability to calculating ligand-protein binding energies. We test this using the PLF547 dataset \cite{plf547}, which has ligands that are significantly larger than the molecules in DES15K and includes chemical groups that are not present in DES15K. The results are shown in Table \ref{tbl:DES15K_PLF547} and Fig. \ref{fig:PLF547}. DensIP achieves an RMSE of 0.7 kcal/mol on neutral protein-ligand fragments showing that it is transferable to larger molecules. Using DFT densities instead of ML ones decreases this error to 0.5 kcal/mol, showing that the DenSNet models have room for improvement with regards to how they generalize to larger molecules. Given the fact that the DenSNet model is only trained on 20,000 calculations, it is quite feasible to generate a larger dataset with larger molecules to improve it's transferability. Finally, given the fact that DensIP achieves similar accuracies to state-of-the-art MLFFs, it may be useful for ranking protein-ligand binding interactions.

Having demonstrated DensIP's transferability and accuracy, we now investigate sources of error in DensIP to guide future development. A significant advantage of a component based model over an MLFF is that individual components of the model can be tested against SAPT component energies to reveal which components can be improved. However, there are a few caveats to this analysis. One is that the energy decomposition is non-unique \cite{morokuma_nonuniqueness}. Another is that it is challenging to compute an energy partitioning method for CCSD(T) energies; SAPT(CC) is expensive \cite{korona2010saptcc}. Instead, we refit DensIP to PBE+MBD energies using the same optimized dimer training dataset and fitting procedure and then compare each component using the following decomposition:
\begin{equation}
\label{eq:E_int_decomp}
 E_{int, PBE+MBD} = E_{elst, SAPT} + E_{exch, SAPT}  + E_{ind} + E_{MBD}, 
\end{equation}
where $E_{ind}$, the induction energy, is
\begin{equation}
\label{eq:E_ind}
E_{ind} =E_{int, PBE} - E_{elst, SAPT} - E_{exch, SAPT}. 
\end{equation}
This decomposition folds the $\delta_{HF}$ term into the induction energy, recovering higher-order induction energies \cite{patkowski2020review}. However, it also includes PBE's delocalization error into the induction energy \cite{cohen2012delocalization, johnson2023delocalization}. Nevertheless, this allows one to have a quantitative assessment of each of the components. It especially allows a rigorous assessment of the effect of errors in the DenSNet-AE densities on the electrostatic and dispersion terms. 

Using this analysis we show the error of each component in Table \ref{tbl:component_errors}. It can be readily seen that the exchange repulsions and induction components are the largest source of errors.  We show in the SI that replacing ML densities with ab-initio densities for the exchange repulsion and induction components does not noticeably reduce these errors in the short-range, so they can be attributed to the functional form of these components \cite{SI}. However, it can be seen that there is an improvement for the medium range-dimers, indicating that the tails of the ML densities do introduce errors in the overlaps at long distances. Indeed, sometimes the overlap integral is slightly negative for distant dimers computed in Fig.\ref{fig:long_range} and in those cases the overlap was set to 0. This introduced a slight error in the model at long-range. Thus, we conclude that both DensIP's exchange repulsion and induction components and the DenSNet model have room for improvement.


\begin{table}[hbpt]
\centering
  \caption{ RMSE of DensIP components on the optimized dimer dataset with respect to PBE+MBD interaction energies decomposed according to Eq. \ref{eq:E_int_decomp} (kcal/mol)}
  \label{tbl:component_errors}
  \begin{tabular}{lcccc}
    \hline
   &   Repulsive wall & Short-range & Equilibrium & Medium-range  \\
    \hline
$E_{\textrm{elst}}$ & 0.5 & 0.4 & 0.2 & 0.1 \\ 
$E_{\textrm{exch}}$ & 3.4 & 2.0 & 0.8 & 0.2 \\
$E_{\textrm{ind}}$ & 2.5 & 1.8 & 0.8 & 0.1 \\ 
$E_{\textrm{disp}}$ & 0.0 & 0.0 & 0.0 & 0.0 \\ 

    \hline

  \end{tabular}
  \end{table}

\begin{figure}[htbp]
    \centering
        \includegraphics[width=.7\textwidth]{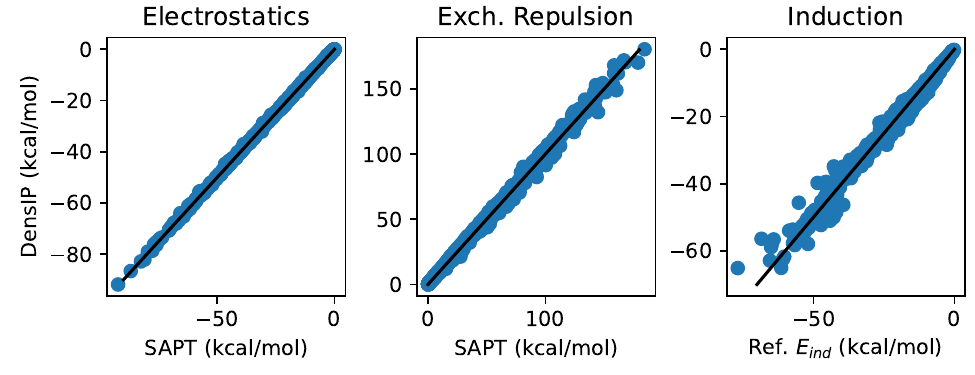}
        \hfill
    \caption{The individual components of DensIP fitted to PBE+MBD and compared to the energy decomposition of PBE+MBD in eq. \ref{eq:E_int_decomp}.}
    \label{fig:components}
\end{figure}

Next, we investigate if DensIP performs equally well for all classes of organic compounds or if some classes are more challenging than others. Looking at Table \ref{tbl:DES15K_groups}, we see that the model performs significantly worse for nitriles and carboxylic acids. This is to be expected as the underlying AVDO exchange repulsion model has difficulties with these systems \cite{avdo}. Nitriles are difficult for the AVDO approximation because the definition of the valence density excludes one of the bonding orbitals in the C-N triple bond. Since the configuration is linear, this bond is accessible in dimer interactions and plays a part in the exchange repulsion interaction. Carboxylic acids present a problem because they often form two hydrogen bonds with another molecule. This causes the dimer to have a shorter distance and thus have higher errors from the AVDO model.

 \begin{table}
  \caption{RMSE of DensIP on the optimized dimer dataset by compound class (kcal/mol)}
  \label{tbl:DES15K_groups}
  \centering
  \begin{tabular}{lcccc}
  \hline
  Model
    & \shortstack{ Repulsive wall }
     & \shortstack{ Short-range }
    & \shortstack{ Equilibrium }
    & \shortstack{ Medium-range } \\
    \hline
alkanes               & 1.3  & 0.9  & 0.3  & 0.1  \\
alkenes               & 2.0  & 1.1  & 0.4  & 0.1  \\
esters                & 2.3  & 1.2  & 0.3  & 0.1  \\
aldehydes             & 2.5  & 1.6  & 0.5  & 0.1  \\
water                 & 2.8  & 1.3  & 0.5  & 0.1  \\
H$_2$                 & 2.8  & 1.6  & 0.6  & 0.2  \\
arenes                & 2.9  & 1.8  & 0.7  & 0.2  \\
alkynes               & 2.9  & 1.4  & 0.5  & 0.1  \\
ethers                & 2.9  & 1.9  & 0.7  & 0.1  \\
alcohols              & 3.0  & 1.6  & 0.6  & 0.2  \\
ketones               & 3.2  & 1.4  & 0.5  & 0.1  \\
amides                & 4.2  & 2.2  & 0.7  & 0.1  \\
N-heteroarenes        & 4.3  & 1.9  & 0.6  & 0.2  \\
amines                & 4.3  & 2.3  & 0.7  & 0.2  \\
carboxylic acids      & 5.3  & 3.6  & 1.2  & 0.3  \\
nitriles              & 6.4  & 2.8  & 0.5  & 0.1  \\
\hline
    \end{tabular}
  \end{table}

To provide context for the developments in DensIP, we compare DensIP to previous density-based intermolecular potentials. IPML was the first density-based intermolecular potential to test transferability to molecules outside of its training dataset \cite{2018ipml}. It has an RMSE of 0.7 kcal/mol on the SSI dataset, indicating that at equilibrium it probably has a similar performance as DensIP. However, it predicts a non-smooth potential energy surface and spurious short-range behavior that causes it to have an RMSE of 4.7 kcal/mol on the S66x8 dataset compared to CCSD(T) \cite{2021cliff}. Thus, it is not appropriate for use in generating synthetic datasets. More recently, CLIFF, a model whose design was inspired by IPML \cite{2021cliff}, has improved smoothness, higher accuracy, and covers more elements, albeit at the cost of more atom-type parameters \cite{2021cliff}. It has an RMSE of 1.1 kcal/mol on S66x8 compared to DensIP's RMSE of 0.6 kcal/mol on the same dataset compared to the SAPT2(+3)$\delta$MP2 interaction energies, which CLIFF was trained on, see Table \ref{tbl:DensIP_CLIFF}. Comparing DensIP to SAPT2+(3)$\delta$MP2 components we see that a large reduction of error with respect to CLIFF occurs in the electrostatic energy, likely due to the better description of the density. It is difficult to make conclusions about the large errors in DensIP's in induction and dispersion due to the fact that SAPT2(+3)$\delta$MP2 induction and dispersion components may not be appropriate reference values because the $\delta$MP2 correction mixes components \cite{2021cliff, schriber2025_benchmark_sapt} and MP2 is generally known to overestimate dispersion \cite{cybulski2007_mp2_overbind, riley2012_mp2_overbind, patkowski2020review}. The fact that the error in the exchange repulsion is larger than the error in the interaction energy suggests that there is some error compensation in DensIP. With regards to transferability, we note that DensIP contains 4 universal parameters compared to the 44 atom-type parameters used in CLIFF. CLIFF was tested only on biologically relevant systems, while here we demonstrate transferability across a broader region of chemical space of molecules containing H, C, N, and O and broadly demonstrate that this extends to non-equilibrium conformations. On the other hand, we note that CLIFF is also able to treat molecules containing F, S, Cl, and Br. Given that both the AVDO exchange-repulsion model and DenSNet have been successfully applied to heavier atoms \cite{avdo, bogojeski2026enhancing}, there is no indication that the DensIP framework is intrinsically restricted to the elements considered here. However, establishing its performance for additional elements requires generating more training data and systematic validation and thus lies beyond the scope of the present study. Finally, we note that comparison with general-purpose MLFFs is what is most important. In order for DensIP to be useful, it must present an improvement over these MLFFs, which we have shown in the case of long-range interactions. In light of the fact that even a general-purpose MLFF with long-range components still performs worse than DensIP despite being trained on 100 million DFT calculations that include molecular clusters that sample long-range interactions, it is possible that a very large dataset will be needed to sample long-range interactions to further improve MLFF models. This is something that DensIP is in a position to provide.

 \begin{table}
  \caption{Comparison of CLIFF and DensIP \\ RMSE (kcal/mol) on the S66x8 dataset, }
  \label{tbl:DensIP_CLIFF}
  \centering
  \begin{tabular}{lc c}
  \hline
  Model
    & CLIFF  & DensIP \\
    \hline
Electrostatics   & 0.8  &  0.3\\
exchange repulsion   & 1.1  & 0.9\\
Induction   & 0.6  &  3.9\\
Dispersion   & 0.3  & 3.2\\
$E_{int}$ SAPT2(+3)$\delta$MP2 & 1.1 & 0.6\\
$E_{int}$  CCSD(F12*)(T) & 1.2  & 0.6\\
\hline
    \end{tabular}
  \end{table}

With regards to computational expense, evaluating the electron density with DenSNet is roughly 3 orders of magnitude cheaper than DFT \cite{bogojeski2026enhancing}. Currently, the slowest step in calculating interaction energies in DensIP is computing electrostatic energies because DenSNet-AE represents the density as $\Delta \rho_{density fit} + \rho_{SAD}$ and the $\rho_{SAD}$-$\rho_{SAD}$ electrostatic energy is a 4-center integral. This limits the calculation time for interaction energies to the order of seconds per configuration per core on a CPU node. For instance, a DFT calculation of a benzene dimer interaction energy using the PBE functional on our hardware \cite{UL_HPC} takes 2,000 cpu seconds whereas DensIP takes 30 cpu seconds. Changing the ML model to represent the entire density only using a density fitting basis will likely lead to an order of magnitude improvement, such that the production of large synthetic datasets begins to be viable. Another important point is that DenSNet is linearly scaling, the electrostatic and overlap calculations formally are quadratically scaling, and the polarization model and MBD are cubically scaling with a small prefactor compared to DFT. Thus, datasets with larger molecules than can currently be calculated with DFT are also possible, though the indication from PLF547 is that further testing with system size is needed.

To summarize the main avenues for improving DensIP:

\begin{itemize}
\item The model could be extended to a broader range of elements by training DenSNet on a more diverse dataset of electron densities such as QCML \cite{2025qcml} or OMOL25 \cite{omol25} and subsequently reparameterizing DensIP.
\item Computational efficiency could be improved by using only a density-fitting basis set to represent the all-electron density in DenSNet-AE.
\item DenSNet could be trained on a dataset containing larger molecular systems and potentially use electron densities computed at a higher level of theory.
\item The exchange repulsion and induction components require further refinement to improve their accuracy in the short-range regime.
\end{itemize}

\section{Summary}
In conclusion, we present DensIP, a 4-parameter intermolecular potential that uses electronic densities from an equivariant ML model, DenSNet, as a central quantity for each of its components. We demonstrate its transferability to new molecules and non-equilibrium conformations of small, neutral, closed-shell organic molecules composed of H, C, N, and O, where it achieves a 0.7 kcal/mol RMSE for near equilibrium configurations including non-equilibrium configurations taken from molecular dynamics trajectories. We further demonstrate DensIP's transferability to molecules much larger than those in its training dataset, specifically to protein-ligand interactions, where it achieves accuracies comparable to state-of-the-art MLFFs, with an RMSE of 0.7 kcal/mol. DensIP is significantly more data efficient than these MLFFs, being trained only on 20,000 DFT calculations and 1,016 CCSD(T) calculations and this means that it is feasible to generate more data to improve DensIP. Importantly, DensIP already exceeds the accuracy of these MLFFs in the long-range, providing near CCSD(T)/CBS accuracy. Given that the method is orders of magnitude less expensive than DFT, it is an attractive avenue for producing high-accuracy synthetic data in the long-range, potentially with larger molecules than those present in existing datasets, that can then be used to further improve general-purpose MLFFs. The Hellmann-Feynman theorem \cite{feynman1939forces} and Hohenberg-Kohn theorem \cite{hohenberg1964inhomogeneous} prove that the electronic density contains all the information needed for an intermolecular potential. Here, we have shown that a simple empirical model can extract such information accurately and transferably from accurate ML densities.

\section*{Associated content}
\section*{Data Availability Statement}
The data underlying this study are openly available on Zenodo at xxx.

\section*{Supporting information}
The Supporting Information is available free of charge at xxxx.


\section*{Acknowledgements}
This research was funded by the Luxembourg National Research Fund (FNR), grant reference MBD-in-BMD C23/MS/18093472, the European Research Council (ERC) Project FITMOL-101054629, and the German Ministry of Research, Technology and Space (BMFTR) under Grants 13GW0744A, 01IS14013A-E, 01GQ1115, 01GQ0850, 01IS18025A, 031L0207D, and 01IS18037A. K.R.M. was partly supported by the Institute of Information \& Communications Technology Planning \& Evaluation (IITP) grants funded by the Korea government (MSIT) (No. 2019-0-00079, Artificial Intelligence Graduate School Program, Korea University and No. 2022-0-00984, Development of Artificial Intelligence Technology for Personalized Plug-and-Play Explanation and Verification of Explanation), DFG and Hector Science Academy. The calculations presented in this paper were carried out using the HPC facilities of the University of Luxembourg \cite{UL_HPC} (see \href{http://hpc.uni.lu}{hpc.uni.lu}).

\printbibliography

@article{2021cliff,
    author = {Schriber, Jeffrey B. and Nascimento, Daniel R. and Koutsoukas, Alexios and Spronk, Steven A. and Cheney, Daniel L. and Sherrill, C. David},
    title = {CLIFF: A component-based, machine-learned, intermolecular force field},
    journal = {The Journal of Chemical Physics},
    volume = {154},
    number = {18},
    pages = {184110},
    year = {2021},
    month = {05},
    issn = {0021-9606},
    doi = {10.1063/5.0042989},
    url = {https://doi.org/10.1063/5.0042989},
}

@article{avdo,
  title={Accurate and Transferable Pauli Exchange-Repulsion for Molecules with the Anisotropic Valence Density Overlap Model},
  author={Wing, Dahvyd and Tkatchenko, Alexandre},
  journal={arXiv preprint arXiv:2510.25629},
  year={2025}
}

@article{2016van_vleet_slater_isa,
  title={Beyond Born--Mayer: Improved models for short-range repulsion in ab initio force fields},
  author={Van Vleet, Mary J and Misquitta, Alston J and Stone, Anthony J and Schmidt, Jordan R},
  journal={Journal of Chemical Theory and Computation},
  volume={12},
  number={8},
  pages={3851--3870},
  year={2016},
  publisher={ACS Publications}
}

@article{2018ipml,
  title={Non-covalent interactions across organic and biological subsets of chemical space: Physics-based potentials parametrized from machine learning},
  author={Bereau, Tristan and DiStasio Jr, Robert A and Tkatchenko, Alexandre and Von Lilienfeld, O Anatole},
  journal={The Journal of Chemical Physics},
  volume={148},
  number={24},
  pages={241706},
  year={2018},
  publisher={AIP Publishing LLC}
}

@article{2024sampl9,
  title={The SAMPL9 host--guest blind challenge: an overview of binding free energy predictive accuracy},
  author={Amezcua, Martin and Setiadi, Jeffry and Mobley, David L},
  journal={Physical Chemistry Chemical Physics},
  volume={26},
  number={12},
  pages={9207--9225},
  year={2024},
  publisher={Royal Society of Chemistry}
}

@article{2010medicinal,
  title={A medicinal chemist’s guide to molecular interactions},
  author={Bissantz, Caterina and Kuhn, Bernd and Stahl, Martin},
  journal={Journal of Medicinal Chemistry},
  volume={53},
  number={14},
  pages={5061--5084},
  year={2010},
  publisher={ACS Publications}
}

@article{2021_csp_review,
  title={Crystal structure prediction methods for organic molecules: State of the art},
  author={Bowskill, David H and Sugden, Isaac J and Konstantinopoulos, Stefanos and Adjiman, Claire S and Pantelides, Constantinos C},
  journal={Annual Review of Chemical and Biomolecular Engineering},
  volume={12},
  pages={593--623},
  year={2021},
  publisher={Annual Reviews}
}

@article{tretiakov2025NCI_review,
    author = {Tretiakov, Serhii and Nigam, AkshatKumar and Pollice, Robert},
    title = {Studying Noncovalent
Interactions in Molecular Systems
with Machine Learning},
    journal = {Chemical Reviews},
    volume = {125},
    number = {12},
    pages = {5776-5829},
    year = {2025},
    month = {06},
    issn = {0009-2665},
    doi = {10.1021/acs.chemrev.4c00893},
    url = {https://doi.org/10.1021/acs.chemrev.4c00893},
}

@article{2025casp16,
author = {Gilson, Michael K. and Eberhardt, Jerome and {\v{S}}krinjar, Peter and Durairaj, Janani and Robin, Xavier and Kryshtafovych, Andriy},
title = {Assessment of Pharmaceutical Protein-Ligand Pose and Affinity Predictions in CASP16},
journal = {Proteins: Structure, Function, and Bioinformatics},
volume = {94},
number = {1},
pages = {249-266},
doi = {https://doi.org/10.1002/prot.70061},
url = {https://onlinelibrary.wiley.com/doi/abs/10.1002/prot.70061},
eprint = {https://onlinelibrary.wiley.com/doi/pdf/10.1002/prot.70061},
year = {2026}
}

@article{2019d3r,
  title={D3R Grand Challenge 3: blind prediction of protein--ligand poses and affinity rankings},
  author={Gaieb, Zied and Parks, Conor D and Chiu, Michael and Yang, Huanwang and Shao, Chenghua and Walters, W Patrick and Lambert, Millard H and Nevins, Neysa and Bembenek, Scott D and Ameriks, Michael K and others},
  journal={Journal of Computer-Aided Molecular Design},
  volume={33},
  number={1},
  pages={1--18},
  year={2019},
  publisher={Springer}
}

@article{mace-omol,
  title={Cross learning between electronic structure theories for unifying molecular, surface, and inorganic crystal foundation force fields},
  author={Batatia, Ilyes and Lin, Chen and Hart, Joseph and Kasoar, Elliott and Elena, Alin M and Norwood, Sam Walton and Wolf, Thomas and Cs{\'a}nyi, G{\'a}bor},
  journal={arXiv preprint arXiv:2510.25380},
  year={2025}
}

@article{vandenbrande2017MEDFF,
  title={The monomer electron density force field (MEDFF): A physically inspired model for noncovalent interactions},
  author={Vandenbrande, Steven and Waroquier, Michel and Speybroeck, Veronique Van and Verstraelen, Toon},
  journal={Journal of Chemical Theory and Computation},
  volume={13},
  number={1},
  pages={161--179},
  year={2017},
  publisher={ACS Publications}
}

@article{hoja2021qm7,
  title={QM7-X, a comprehensive dataset of quantum-mechanical properties spanning the chemical space of small organic molecules},
  author={Hoja, Johannes and Medrano Sandonas, Leonardo and Ernst, Brian G and Vazquez-Mayagoitia, Alvaro and DiStasio Jr, Robert A and Tkatchenko, Alexandre},
  journal={Scientific Data},
  volume={8},
  number={1},
  pages={43},
  year={2021},
  publisher={Nature Publishing Group UK London}
}

@article{frank2022so3krates,
  title={So3krates: Equivariant attention for interactions on arbitrary length-scales in molecular systems},
  author={Frank, Thorben and Unke, Oliver and M{\"u}ller, Klaus-Robert},
  journal={Advances in Neural Information Processing Systems},
  volume={35},
  pages={29400--29413},
  year={2022}
}

@article{hohenberg1964inhomogeneous,
  title={Inhomogeneous electron gas},
  author={Hohenberg, P and Kohn, W},
  journal={Physical Review},
  volume={136},
  number={3B},
  pages={B864--B871},
  year={1964},
  doi={10.1103/PhysRev.136.B864}
}

@article{rupp2012fast,
  title={Fast and accurate modeling of molecular atomization energies with machine learning},
  author={Rupp, Matthias and Tkatchenko, Alexandre and M{\"u}ller, Klaus-Robert and von Lilienfeld, O. Anatole},
  journal={Physical Review Letters},
  volume={108},
  number={5},
  pages={058301},
  year={2012},
  doi={10.1103/PhysRevLett.108.058301}
}

@article{snyder2012finding,
  title={Finding density functionals with machine learning},
  author={Snyder, John C and Rupp, Matthias and Hansen, Katja and M{\"u}ller, Klaus-Robert and Burke, Kieron},
  journal={Physical Review Letters},
  volume={108},
  number={25},
  pages={253002},
  year={2012},
  doi={10.1103/PhysRevLett.108.253002}
}

@article{li2014understanding,
  title={Understanding machine-learned density functionals},
  author={Li, Li and Snyder, John C and Pelaschier, Isabelle M and Huang, Jessica and Niranjan, Uma-Naresh and Duncan, Paul and Rupp, Matthias and M{\"u}ller, Klaus-Robert and Burke, Kieron},
  journal={International Journal of Quantum Chemistry},
  volume={116},
  pages={819--833},
  year={2016},
  doi={10.1002/qua.25040}
}

@article{brockherde2017bypassing,
  title={Bypassing the {K}ohn--{S}ham equations with machine learning},
  author={Brockherde, Felix and Vogt, Leslie and Li, Li and Tuckerman, Mark E and Burke, Kieron and M{\"u}ller, Klaus-Robert},
  journal={Nature Communications},
  volume={8},
  pages={872},
  year={2017},
  doi={10.1038/s41467-017-00839-3}
}

@article{bogojeski2020quantum,
  title={Quantum chemical accuracy from density functional approximations via machine learning},
  author={Bogojeski, Mihail and Vogt-Maranto, Leslie and Tuckerman, Mark E and M{\"u}ller, Klaus-Robert and Burke, Kieron},
  journal={Nature Communications},
  volume={11},
  pages={5223},
  year={2020},
  doi={10.1038/s41467-020-19093-1}
}

@article{bai2022machine,
  title={Machine learning the {H}ohenberg--{K}ohn map for molecular excited states},
  author={Bai, Yuanming and Vogt-Maranto, Leslie and Tuckerman, Mark E and Glover, William J},
  journal={Nature Communications},
  volume={13},
  pages={7044},
  year={2022},
  doi={10.1038/s41467-022-34436-w}
}

@article{shao2023machine,
  title={Machine learning electronic structure methods based on the one-electron reduced density matrix},
  author={Shao, Xuecheng and Paetow, Lukas and Tuckerman, Mark E and Pavanello, Michele},
  journal={Nature Communications},
  volume={14},
  year={2023},
  pages={6281},
  doi={10.1038/s41467-023-41953-9}
}

@article{kirkpatrick2021pushing,
  title={Pushing the frontiers of density functionals by solving the fractional electron problem},
  author={Kirkpatrick, James and McMorrow, Brendan and Turban, David H P and Gaunt, Alexander L and Spencer, James S and Matthews, Alexander G D G and Obika, Annette and Thiry, Louis and Fortunato, Meire and Pfau, David and others},
  journal={Science},
  volume={374},
  number={6573},
  pages={1385--1389},
  year={2021},
  doi={10.1126/science.abj6511}
}

@article{huang2023dft,
  title={The central role of density functional theory in the {AI} age},
  author={Huang, Bing and von Rudorff, Guido and von Lilienfeld, O. Anatole},
  journal={Science},
  volume={381},
  number={6655},
  pages={170--175},
  year={2023},
  doi={10.1126/science.abn3445}
}

@inproceedings{
kaniselvan2025helm,
title={Learning from the Electronic Structure of Molecules across the Periodic Table},
author={Manasa Kaniselvan and Benjamin Kurt Miller and Meng Gao and Juno Nam and Daniel S. Levine},
booktitle={The Fourteenth International Conference on Learning Representations},
year={2026},
url={https://openreview.net/forum?id=PS1YS8Wv4t}
}

@article{khan2025adapting,
  title={Adapting hybrid density functionals with machine learning},
  author={Khan, Danish and Price, Alastair and Huang, Bing and Ach, Maximilian L and von Lilienfeld, O. Anatole},
  journal={Science Advances},
  volume={11},
  number={5},
  pages={eadt7769},
  year={2025},
  doi={10.1126/sciadv.adt7769}
}

@misc{thomas2018tensor,
      title={Tensor field networks: Rotation- and translation-equivariant neural networks for 3D point clouds}, 
      author={Nathaniel Thomas and Tess Smidt and Steven Kearnes and Lusann Yang and Li Li and Kai Kohlhoff and Patrick Riley},
      year={2018},
      eprint={1802.08219},
      archivePrefix={arXiv},
      primaryClass={cs.LG},
      url={https://arxiv.org/abs/1802.08219}, 
}

@inproceedings{weiler20183d,
  title={3D steerable CNNs: Learning rotationally equivariant features in volumetric data},
  author={Weiler, Maurice and Geiger, Mario and Welling, Max and Boomsma, Wouter and Cohen, Taco},
  booktitle={Advances in Neural Information Processing Systems},
  volume={31},
  year={2018},
  eprint={1807.02547},
  archiveprefix={arXiv}
}

@inproceedings{unke2021se,
  title={SE(3)-equivariant prediction of molecular wavefunctions and electronic densities},
  author={Unke, Oliver and Bogojeski, Mihail and Gastegger, Michael and Geiger, Mario and Smidt, Tess and M{\"u}ller, Klaus-Robert},
  booktitle={Advances in Neural Information Processing Systems},
  volume={34},
  pages={14434--14447},
  year={2021},
  eprint={2106.02347},
  archiveprefix={arXiv}
}

@article{grisafi2018transferable,
  title={Transferable machine-learning model of the electron density},
  author={Grisafi, Andrea and Wilkins, David M and Meyer, Benjamin A R and Fabrizio, Alberto and Corminboeuf, Cl{\'e}mence and Ceriotti, Michele},
  journal={ACS Central Science},
  volume={5},
  number={1},
  pages={57--64},
  year={2019},
  doi={10.1021/acscentsci.8b00551}
}

@article{grisafi2022electronic,
  title={Electronic-structure properties from atom-centered predictions of the electron density},
  author={Grisafi, Andrea and Lewis, Alan M and Rossi, Mariana and Ceriotti, Michele},
  journal={Journal of Chemical Theory and Computation},
  volume={19},
  pages={3509--3525},
  year={2023},
  doi={10.1021/acs.jctc.2c00850}
}

@article{rackers2022cracking,
  title={A recipe for cracking the quantum scaling limit with machine learned electron densities},
  author={Rackers, Joshua A and Tecot, Lucas and Geiger, Mario and Smidt, Tess E},
  journal={Machine Learning: Science and Technology},
  volume={4},
  year={2023},
  pages={015027},
  doi={10.1088/2632-2153/acb314},
  eprint={2201.03726},
  archiveprefix={arXiv}
}

@phdthesis{bogojeski2023machine,
  title={Machine learning for electronic structure},
  author={Bogojeski, Mihail},
  year={2023},
  school={Technische Universit{\"a}t Berlin}
}

@inproceedings{elsborg2026electra,
 author = {Elsborg, Jonas and Thiede, Luca and Aspuru-Guzik, Alan and Vegge, Tejs and Bhowmik, Arghya},
 booktitle = {Advances in Neural Information Processing Systems},
 doi = {10.52202/085713-0947},
 editor = {D. Belgrave and C. Zhang and H. Lin and R. Pascanu and P. Koniusz and M. Ghassemi and N. Chen},
 pages = {28092--28121},
 publisher = {Curran Associates, Inc.},
 title = {ELECTRA: A Cartesian Network for 3D Charge Density Prediction with Floating Orbitals},
 url = {https://proceedings.neurips.cc/paper_files/paper/2025/file/288b63aa98084366c4536ba0574a0f22-Paper-Conference.pdf},
 volume = {38, Main Conference},
 year = {2025}
}

@article{jorgensen2022equivariant,
  title={Equivariant graph neural networks for fast electron density estimation of molecules, liquids, and solids},
  author={J{\o}rgensen, Peter Bj{\o}rn and Bhowmik, Arghya},
  journal={npj Computational Materials},
  volume={8},
  number={1},
  pages={183},
  year={2022},
  publisher={Nature Publishing Group UK London}
}

@article{koker2024higher,
  title={Higher-order equivariant neural networks for charge density prediction in materials},
  author={Koker, Teddy and Quigley, Keegan and Taw, Eric and Tibbetts, Kevin and Li, Lin},
  journal={npj Computational Materials},
  volume={10},
  number={1},
  pages={161},
  year={2024},
  publisher={Nature Publishing Group UK London}
}

@article{li2024superres,
  title={Image super-resolution inspired electron density prediction},
  author={Li, Chenghan and Sharir, Or and Yuan, Shunyue and Chan, Garnet Kin-Lic},
  journal={Nature Communications},
  volume={16},
  year={2025},
  pages={4811},
  doi={10.1038/s41467-025-60095-8},
  eprint={2402.12335},
  archiveprefix={arXiv}
}

@article{fu2024recipe,
  title={A recipe for charge density prediction},
  author={Fu, Xiang and Rosen, Andrew and Bystrom, Kyle and Wang, Rui and Musaelian, Albert and Kozinsky, Boris and Smidt, Tess and Jaakkola, Tommi},
  journal={Advances in Neural Information Processing Systems},
  volume={37},
  pages={9727--9752},
  year={2024}
}

@article{bogojeski2026enhancing,
  title={Enhancing molecular dynamics with equivariant machine-learned densities},
  author={Bogojeski, Mihail and Hasyim, Muhammad R and Vogt-Maranto, Leslie and M{\"u}ller, Klaus-Robert and Burke, Kieron and Tuckerman, Mark E},
  journal={arXiv preprint arXiv:2604.24563},
  year={2026}
}

@article{kingma2014adam,
  title={Adam: A method for stochastic optimization},
  author={Kingma, Diederik P and Ba, Jimmy},
  journal={arXiv preprint arXiv:1412.6980},
  year={2014}
}

@article{stone_2009_CT,
title = {Charge-transfer in Symmetry-Adapted Perturbation Theory},
journal = {Chemical Physics Letters},
volume = {473},
number = {1},
pages = {201-205},
year = {2009},
issn = {0009-2614},
doi = {https://doi.org/10.1016/j.cplett.2009.03.073},
url = {https://www.sciencedirect.com/science/article/pii/S0009261409003947},
author = {Anthony J. Stone and Alston J. Misquitta}
}

@article{elking2010gmm,
  title={Gaussian multipole model (GMM)},
  author={Elking, Dennis M and Cisneros, G Andr{\'e}s and Piquemal, Jean-Philip and Darden, Thomas A and Pedersen, Lee G},
  journal={Journal of Chemical Theory and Computation},
  volume={6},
  number={1},
  pages={190--202},
  year={2010},
  publisher={ACS Publications}
}

@article{elking_2007_pol,
  title={Gaussian induced dipole polarization model},
  author={Elking, Dennis and Darden, TOM and Woods, Robert J},
  journal={Journal of Computational Chemistry},
  volume={28},
  number={7},
  pages={1261--1274},
  year={2007},
  publisher={Wiley Online Library}
}

@article{so3lr,
  title={Molecular simulations with a pretrained neural network and universal pairwise force fields},
  author={Kabylda, Adil and Frank, J Thorben and Su{\'a}rez-Dou, Sergio and Khabibrakhmanov, Almaz and Medrano Sandonas, Leonardo and Unke, Oliver T and Chmiela, Stefan and Müller, Klaus-Robert and Tkatchenko, Alexandre},
  journal={Journal of the American Chemical Society},
  volume={147},
  number={37},
  pages={33723--33734},
  year={2025},
  publisher={ACS Publications}
}

@article{mace-off,
  title={Mace-off: Short-range transferable machine learning force fields for organic molecules},
  author={Kov{\'a}cs, D{\'a}vid P{\'e}ter and Moore, J Harry and Browning, Nicholas J and Batatia, Ilyes and Horton, Joshua T and Pu, Yixuan and Kapil, Venkat and Witt, William C and Magdau, Ioan-Bogdan and Cole, Daniel J and others},
  journal={Journal of the American Chemical Society},
  volume={147},
  number={21},
  pages={17598--17611},
  year={2025},
  publisher={ACS Publications}
}

@article{mace-polar,
  title={MACE-POLAR-1: A polarisable electrostatic foundation model for molecular chemistry},
  author={Batatia, Ilyes and Baldwin, William J and Kuryla, Domantas and Hart, Joseph and Kasoar, Elliott and Elena, Alin M and Moore, Harry and Gawkowski, Miko{\l}aj J and Shi, Benjamin X and Kapil, Venkat and others},
  journal={arXiv preprint arXiv:2602.19411},
  year={2026}
}

@book{stone_2013_book,
  title={The Theory of Intermolecular Forces},
  author={Stone, Anthony},
  year={2013},
  publisher={Oxford University Press}
}

@article{unke2021mlff,
  title={Machine learning force fields},
  author={Unke, Oliver T and Chmiela, Stefan and Sauceda, Huziel E and Gastegger, Michael and Poltavsky, Igor and Schutt, Kristof T and Tkatchenko, Alexandre and M{\"u}ller, Klaus-Robert},
  journal={Chemical Reviews},
  volume={121},
  number={16},
  pages={10142--10186},
  year={2021},
  publisher={ACS Publications}
}

@article{uma,
  title={UMA: A family of universal models for atoms},
  author={Wood, Brandon and Dzamba, Misko and Fu, Xiang and Gao, Meng and Shuaibi, Muhammed and Barroso-Luque, Luis and Abdelmaqsoud, Kareem and Gharakhanyan, Vahe and Kitchin, John and Levine, Daniel and others},
  journal={Advances in Neural Information Processing Systems},
  volume={38},
  pages={129391--129427},
  year={2026}
}

@article{ccsdt,
  title={Coupled-cluster theory in quantum chemistry},
  author={Bartlett, Rodney J and Musia{\l}, Monika},
  journal={Reviews of Modern Physics},
  volume={79},
  number={1},
  pages={291--352},
  year={2007},
  publisher={APS}
}

@book{martin_electronic_structure_book,
  title={Electronic structure: basic theory and practical methods},
  author={Martin, Richard M},
  year={2020},
  publisher={Cambridge university press}
}

@inproceedings{parr_and_yang,
  title={Density functional theory of atoms and molecules},
  author={Parr, Robert G},
  booktitle={Horizons of Quantum Chemistry: Proceedings of the Third International Congress of Quantum Chemistry Held at Kyoto, Japan, October 29-November 3, 1979},
  pages={5--15},
  year={1989},
  organization={Springer}
}

@article{des370k,
  title={Quantum chemical benchmark databases of gold-standard dimer interaction energies},
  author={Donchev, Alexander G and Taube, Andrew G and Decolvenaere, Elizabeth and Hargus, Cory and McGibbon, Robert T and Law, Ka-Hei and Gregersen, Brent A and Li, Je-Luen and Palmo, Kim and Siva, Karthik and others},
  journal={Scientific Data},
  volume={8},
  number={1},
  pages={55},
  year={2021},
  publisher={Nature Publishing Group UK London}
}

@article{mace_mp,
    author = {Batatia, Ilyes and Benner, Philipp and Chiang, Yuan and Elena, Alin M. and Kov{\a}cs, D{\'a}vid P. and Riebesell, Janosh and Advincula, Xavier R. and Asta, Mark and Avaylon, Matthew and Baldwin, William J. and Berger, Fabian and Bernstein, Noam and Bhowmik, Arghya and Bigi, Filippo and Blau, Samuel M. and C{\ua}rare, Vlad and Ceriotti, Michele and Chong, Sanggyu and Darby, James P. and De, Sandip and Della Pia, Flaviano and Deringer, Volker L. and Elijo{\v}ius, Rokas and El-Machachi, Zakariya and Fako, Edvin and Falcioni, Fabio and Ferrari, Andrea C. and Gardner, John L. A. and Gawkowski, Miko{\l}aj J. and Genreith-Schriever, Annalena and George, Janine and Goodall, Rhys E. A. and Grandel, Jonas and Grey, Clare P. and Grigorev, Petr and Han, Shuang and Handley, Will and Heenen, Hendrik H. and Hermansson, Kersti and Ho, Cheuk Hin and Hofmann, Stephan and Holm, Christian and Jaafar, Jad and Jakob, Konstantin S. and Jung, Hyunwook and Kapil, Venkat and Kaplan, Aaron D. and Karimitari, Nima and Kermode, James R. and Kourtis, Panagiotis and Kroupa, Namu and Kullgren, Jolla and Kuner, Matthew C. and Kuryla, Domantas and Liepuoniute, Guoda and Lin, Chen and Margraf, Johannes T. and Magd{\ua}u, Ioan-Bogdan and Michaelides, Angelos and Moore, J. Harry and Naik, Aakash A. and Niblett, Samuel P. and Norwood, Sam Walton and O’Neill, Niamh and Ortner, Christoph and Persson, Kristin A. and Reuter, Karsten and Rosen, Andrew S. and Rosset, Louise A. M. and Schaaf, Lars L. and Schran, Christoph and Shi, Benjamin X. and Sivonxay, Eric and Stenczel, Tam{\'a}s K. and Sutton, Christopher and Svahn, Viktor and Swinburne, Thomas D. and Tilly, Jules and van der Oord, Cas and Vargas, Santiago and Varga-Umbrich, Eszter and Vegge, Tejs and Vondr{\'a}k, Martin and Wang, Yangshuai and Witt, William C. and Wolf, Thomas and Zills, Fabian and Cs{\'a}nyi, G{\'a}bor},
    title = {A foundation model for atomistic materials chemistry},
    journal = {The Journal of Chemical Physics},
    volume = {163},
    number = {18},
    pages = {184110},
    year = {2025},
    month = {11},
    issn = {0021-9606},
    doi = {10.1063/5.0297006},
    url = {https://doi.org/10.1063/5.0297006},
}

@article{ani1ccx,
  title={Approaching coupled cluster accuracy with a general-purpose neural network potential through transfer learning},
  author={Smith, Justin S and Nebgen, Benjamin T and Zubatyuk, Roman and Lubbers, Nicholas and Devereux, Christian and Barros, Kipton and Tretiak, Sergei and Isayev, Olexandr and Roitberg, Adrian E},
  journal={Nature Communications},
  volume={10},
  number={1},
  pages={2903},
  year={2019},
  publisher={Nature Publishing Group UK London}
}

@article{plf547,
  title={Benchmarking of semiempirical quantum-mechanical methods on systems relevant to computer-aided drug design},
  author={Kriz, Kristian and Rezac, Jan},
  journal={Journal of Chemical Information and Modeling},
  volume={60},
  number={3},
  pages={1453--1460},
  year={2020},
  publisher={ACS Publications}
}

@article{transfer_learning,
  title={Transfer learning for chemically accurate interatomic neural network potentials},
  author={Zaverkin, Viktor and Holzm{\"u}ller, David and Bonfirraro, Luca and K{\"a}stner, Johannes},
  journal={Physical Chemistry Chemical Physics},
  volume={25},
  number={7},
  pages={5383--5396},
  year={2023},
  publisher={The Royal Society of Chemistry}
}

@article{delta_ML_water_michaelides_2025,
  title={Towards Routine Condensed Phase Simulations with Delta-Learned Coupled Cluster Accuracy: Application to Liquid Water},
  author={O’Neill, Niamh and Shi, Benjamin X and Baldwin, William J and Witt, William C and Cs{\'a}nyi, G{\'a}bor and Gale, Julian D and Michaelides, Angelos and Schran, Christoph},
  journal={Journal of Chemical Theory and Computation},
  volume={21},
  number={22},
  pages={11710--11720},
  year={2025},
  publisher={ACS Publications}
}

@article{bowman2022delta,
  title={$\Delta$-machine learned potential energy surfaces and force fields},
  author={Bowman, Joel M and Qu, Chen and Conte, Riccardo and Nandi, Apurba and Houston, Paul L and Yu, Qi},
  journal={Journal of Chemical Theory and Computation},
  volume={19},
  number={1},
  pages={1--17},
  year={2022},
  publisher={ACS Publications}
}

@article{mbis,
  title={Minimal basis iterative stockholder: atoms in molecules for force-field development},
  author={Verstraelen, Toon and Vandenbrande, Steven and Heidar-Zadeh, Farnaz and Vanduyfhuys, Louis and Van Speybroeck, Veronique and Waroquier, Michel and Ayers, Paul W},
  journal={Journal of Chemical Theory and Computation},
  volume={12},
  number={8},
  pages={3894--3912},
  year={2016},
  publisher={ACS Publications}
}

@article{van_vleet2018mastiff,
  title={New angles on standard force fields: Toward a general approach for treating atomic-level anisotropy},
  author={Van Vleet, Mary J and Misquitta, Alston J and Schmidt, JR},
  journal={Journal of Chemical Theory and Computation},
  volume={14},
  number={2},
  pages={739--758},
  year={2018},
  publisher={ACS Publications}
}

@article{xu2018_fragment_method_review,
    author = {Xu, Peng and Guidez, Emilie B. and Bertoni, Colleen and Gordon, Mark S.},
    title = {Perspective: Ab initio force field methods derived from quantum mechanics},
    journal = {The Journal of Chemical Physics},
    volume = {148},
    number = {9},
    pages = {090901},
    year = {2018},
    month = {03},
    issn = {0021-9606},
    doi = {10.1063/1.5009551},
    url = {https://doi.org/10.1063/1.5009551},
}

@article{gem,
    author = {Naseem-Khan, Sehr and Piquemal, Jean-Philip and Cisneros, G. Andr{\'e}s},
    title = {Improvement of the Gaussian Electrostatic Model by separate fitting of Coulomb and exchange-repulsion densities and implementation of a new dispersion term},
    journal = {The Journal of Chemical Physics},
    volume = {155},
    number = {19},
    pages = {194103},
    year = {2021},
    month = {11},
    issn = {0021-9606},
    doi = {10.1063/5.0072380},
    url = {https://doi.org/10.1063/5.0072380},
}

@article{qmdff,
  title={A general quantum mechanically derived force field (QMDFF) for molecules and condensed phase simulations},
  author={Grimme, Stefan},
  journal={Journal of Chemical Theory and Computation},
  volume={10},
  number={10},
  pages={4497--4514},
  year={2014},
  publisher={ACS Publications}
}

@article{fflux,
  title={Description of potential energy surfaces of molecules using FFLUX machine learning models},
  author={Hughes, Zak E and Thacker, Joseph CR and Wilson, Alex L and Popelier, Paul LA},
  journal={Journal of Chemical Theory and Computation},
  volume={15},
  number={1},
  pages={116--126},
  year={2018},
  publisher={ACS Publications}
}

@article{cole2016,
title= {Biomolecular force field parameterization via atoms-in-molecule electron density partitioning},
author={Cole, Daniel J and Vilseck, Jonah Z and Tirado-Rives, Julian and Payne, Mike C and Jorgensen, William L},
journal={Journal of Chemical Theory and Computation},
volume={12},
number={5},
pages={2312--2323},
year={2016},
publisher={ACS Publications}
}

@article{jeziorski1994sapt_review,
  title={Perturbation theory approach to intermolecular potential energy surfaces of van der Waals complexes},
  author={Jeziorski, Bogumil and Moszynski, Robert and Szalewicz, Krzysztof},
  journal={Chemical Reviews},
  volume={94},
  number={7},
  pages={1887--1930},
  year={1994},
  publisher={ACS Publications}
}

@article{thole1981,
  title={Molecular polarizabilities calculated with a modified dipole interaction},
  author={Thole, B Th},
  journal={Chemical Physics},
  volume={59},
  number={3},
  pages={341--350},
  year={1981},
  publisher={Elsevier}
}

@article{TS_rescale_2009,
  title={Accurate molecular van der waals interactions from ground-state electron density and free-atom reference data},
  author={Tkatchenko, Alexandre and Scheffler, Matthias},
  journal={Physical Review Letters},
  volume={102},
  number={7},
  pages={073005},
  year={2009},
  publisher={APS}
}

@article{morokuma_nonuniqueness,
  title={Why do molecules interact? The origin of electron donor-acceptor complexes, hydrogen bonding and proton affinity},
  author={Morokuma, Keiji},
  journal={Accounts of Chemical Research},
  volume={10},
  number={8},
  pages={294--300},
  year={1977},
  publisher={ACS Publications}
}

@article{schriber2025_benchmark_sapt,
  title={Levels of symmetry-adapted perturbation theory (SAPT). II. Convergence of interaction energy components},
  author={Schriber, Jeffrey B and Wallace, Austin M and Cheney, Daniel L and Sherrill, C David},
  journal={The Journal of Chemical Physics},
  volume={163},
  number={8},
  pages={084114},
  year={2025},
  publisher={AIP Publishing}
}

@article{riley2012_mp2_overbind,
  title={Assessment of the Performance of MP2 and MP2 Variants for the Treatment of Noncovalent Interactions},
  author={Riley, Kevin E and Platts, James A and Rezac, Jan and Hobza, Pavel and Hill, J Grant},
  journal={The Journal of Physical Chemistry A},
  volume={116},
  number={16},
  pages={4159--4169},
  year={2012},
  publisher={ACS Publications}
}

@article{cybulski2007_mp2_overbind,
  title={The origin of deficiency of the supermolecule second-order M{\o}ller-Plesset approach for evaluating interaction energies},
  author={Cybulski, Slawomir M and Lytle, Marion L},
  journal={The Journal of Chemical Physics},
  volume={127},
  number={14},
  year={2007},
  pages={141102},
  publisher={AIP Publishing}
}

@article{skylaris2000density_fitting,
  title={On the resolution of identity Coulomb energy approximation in density functional theory},
  author={Skylaris, C-K and Gagliardi, Laura and Handy, Nicholas C and Ioannou, Andrew G and Spencer, Steven and Willetts, Andrew},
  journal={Journal of Molecular Structure: THEOCHEM},
  volume={501},
  pages={229--239},
  year={2000},
  publisher={Elsevier}
}

@article{baerends1973_density_fitting,
  title={Self-consistent molecular Hartree—Fock—Slater calculations I. The computational procedure},
  author={Baerends, E J and Ellis, D E and Ros, P},
  journal={Chemical Physics},
  volume={2},
  number={1},
  pages={41--51},
  year={1973},
  publisher={Elsevier}
}

@article{dunlap1979_density_fitting,
  title={On first-row diatomic molecules and local density models},
  author={Dunlap, BI and Connolly, JWD and Sabin, JR},
  journal={The Journal of Chemical Physics},
  volume={71},
  number={12},
  pages={4993--4999},
  year={1979},
  publisher={American Institute of Physics}
}

@article{weigend2002_jkfit,
  title={A fully direct RI-HF algorithm: Implementation, optimised auxiliary basis sets, demonstration of accuracy and efficiency},
  author={Weigend, Florian},
  journal={Physical Chemistry Chemical Physics},
  volume={4},
  number={18},
  pages={4285--4291},
  year={2002},
  publisher={The Royal Society of Chemistry}
}

@article{hait2018_atomic_polarizability,
  title={How accurate are static polarizability predictions from density functional theory? An assessment over 132 species at equilibrium geometry},
  author={Hait, Diptarka and Head-Gordon, Martin},
  journal={Physical Chemistry Chemical Physics},
  volume={20},
  number={30},
  pages={19800--19810},
  year={2018},
  publisher={The Royal Society of Chemistry}
}

@article{schwerdtfeger2019_atomic_polarizability,
  title={2018 Table of static dipole polarizabilities of the neutral elements in the periodic table},
  author={Schwerdtfeger, Peter and Nagle, Jeffrey K},
  journal={Molecular Physics},
  volume={117},
  number={9-12},
  pages={1200--1225},
  year={2019},
  publisher={Taylor \& Francis}
}

@article{ambrosetti2014_rsscs,
  title={Long-range correlation energy calculated from coupled atomic response functions},
  author={Ambrosetti, Alberto and Reilly, Anthony M and DiStasio, Robert A and Tkatchenko, Alexandre},
  journal={The Journal of Chemical Physics},
  volume={140},
  number={18},
  pages={18A508},
  year={2014},
  publisher={AIP Publishing}
}

@article{mbd,
  title={Accurate and efficient method for many-body van der Waals interactions},
  author={Tkatchenko, Alexandre and DiStasio Jr, Robert A and Car, Roberto and Scheffler, Matthias},
  journal={Physical Review Letters},
  volume={108},
  number={23},
  pages={236402},
  year={2012},
  publisher={APS}
}

@article{nickerson2023_dispersion_benchmark,
  title={Comparison of density-functional theory dispersion corrections for the DES15K database},
  author={Nickerson, Cameron J and Bryenton, Kyle R and Price, Alastair JA and Johnson, Erin R},
  journal={The Journal of Physical Chemistry A},
  volume={127},
  number={41},
  pages={8712--8722},
  year={2023},
  publisher={ACS Publications}
}

@article{libmbd,
  title={libMBD: A general-purpose package for scalable quantum many-body dispersion calculations},
  author={Hermann, Jan and St{\"o}hr, Martin and G{\'o}ger, Szabolcs and Chaudhuri, Shayantan and Aradi, B{\'a}lint and Maurer, Reinhard J and Tkatchenko, Alexandre},
  journal={The Journal of Chemical Physics},
  volume={159},
  number={17},
  pages={174802},
  year={2023},
  publisher={AIP Publishing}
}

@article{2020pyscf,
  title={Recent developments in the PySCF program package},
  author={Sun, Qiming and Zhang, Xing and Banerjee, Samragni and Bao, Peng and Barbry, Marc and Blunt, Nick S and Bogdanov, Nikolay A and Booth, George H and Chen, Jia and Cui, Zhi-Hao and others},
  journal={The Journal of Chemical Physics},
  volume={153},
  number={2},
  year={2020},
  publisher={AIP Publishing}
}

@article{2018pyscf,
  title={PySCF: the Python-based simulations of chemistry framework},
  author={Sun, Qiming and Berkelbach, Timothy C and Blunt, Nick S and Booth, George H and Guo, Sheng and Li, Zhendong and Liu, Junzi and McClain, James D and Sayfutyarova, Elvira R and Sharma, Sandeep and others},
  journal={Wiley Interdisciplinary Reviews: Computational Molecular Science},
  volume={8},
  number={1},
  pages={e1340},
  year={2018},
  publisher={Wiley Online Library}
}

@article{2015libcint,
  title={Libcint: An efficient general integral library for gaussian basis functions},
  author={Sun, Qiming},
  journal={Journal of Computational Chemistry},
  volume={36},
  number={22},
  pages={1664--1671},
  year={2015},
  publisher={Wiley Online Library}
}

@ARTICLE{2020SciPy,
  author  = {Virtanen, Pauli and Gommers, Ralf and Oliphant, Travis E. and
            Haberland, Matt and Reddy, Tyler and Cournapeau, David and
            Burovski, Evgeni and Peterson, Pearu and Weckesser, Warren and
            Bright, Jonathan and {van der Walt}, St{\'e}fan J. and
            Brett, Matthew and Wilson, Joshua and Millman, K. Jarrod and
            Mayorov, Nikolay and Nelson, Andrew R. J. and Jones, Eric and
            Kern, Robert and Larson, Eric and Carey, C J and
            Polat, {\.I}lhan and Feng, Yu and Moore, Eric W. and
            {VanderPlas}, Jake and Laxalde, Denis and Perktold, Josef and
            Cimrman, Robert and Henriksen, Ian and Quintero, E. A. and
            Harris, Charles R. and Archibald, Anne M. and
            Ribeiro, Ant{\^o}nio H. and Pedregosa, Fabian and
            {van Mulbregt}, Paul and {SciPy 1.0 Contributors}},
  title   = {{{SciPy} 1.0: Fundamental Algorithms for Scientific
            Computing in Python}},
  journal = {Nature Methods},
  year    = {2020},
  volume  = {17},
  pages   = {261--272},
  adsurl  = {https://rdcu.be/b08Wh},
  doi     = {10.1038/s41592-019-0686-2},
}

@article{2021qchem,
  title={Software for the frontiers of quantum chemistry: An overview of developments in the Q-Chem 5 package},
  author={Epifanovsky, Evgeny and Gilbert, Andrew TB and Feng, Xintian and Lee, Joonho and Mao, Yuezhi and Mardirossian, Narbe and Pokhilko, Pavel and White, Alec F and Coons, Marc P and Dempwolff, Adrian L and others},
  journal={The Journal of Chemical Physics},
  volume={155},
  number={8},
  pages={084801},
  year={2021},
  publisher={AIP Publishing}
}

@article{rezac2015_charge_transfer_cdft,
  title={Robust, basis-set independent method for the evaluation of charge-transfer energy in noncovalent complexes},
  author={Rezac, Jan and de la Lande, Aurelien},
  journal={Journal of Chemical Theory and Computation},
  volume={11},
  number={2},
  pages={528--537},
  year={2015},
  publisher={ACS Publications}
}

@article{misquitta2013_regularized_sapt_CT,
  title={Charge transfer from regularized symmetry-adapted perturbation theory},
  author={Misquitta, Alston J},
  journal={Journal of Chemical Theory and Computation},
  volume={9},
  number={12},
  pages={5313--5326},
  year={2013},
  publisher={ACS Publications}
}

@article{mao2018_compare_CT,
  title={On the computational characterization of charge-transfer effects in noncovalently bound molecular complexes},
  author={Mao, Yuezhi and Ge, Qinghui and Horn, Paul R and Head-Gordon, Martin},
  journal={Journal of Chemical Theory and Computation},
  volume={14},
  number={5},
  pages={2401--2417},
  year={2018},
  publisher={ACS Publications}
}

@article{2016_almo_CT,
  title={Probing non-covalent interactions with a second generation energy decomposition analysis using absolutely localized molecular orbitals},
  author={Horn, Paul R and Mao, Yuezhi and Head-Gordon, Martin},
  journal={Physical Chemistry Chemical Physics},
  volume={18},
  number={33},
  pages={23067--23079},
  year={2016},
  publisher={The Royal Society of Chemistry}
}

@article{kita1976,
  title={Repulsive potentials for Cl---R and Br---R (R= He, Ne, and Ar) derived from beam experiments},
  author={Kita, S and Noda, K and Inouye, H},
  journal={The Journal of Chemical Physics},
  volume={64},
  number={8},
  pages={3446--3449},
  year={1976},
  publisher={American Institute of Physics}
}

@article{piquemal_darden_2006,
  title={Towards a force field based on density fitting},
  author={Piquemal, Jean-Philip and Cisneros, G Andr{\'e}s and Reinhardt, Peter and Gresh, Nohad and Darden, Thomas A},
  journal={The Journal of Chemical Physics},
  volume={124},
  number={10},
  pages={104101},
  year={2006},
  publisher={AIP Publishing}
}

@article{kim_1981_ovlp,
  title={Dependence of the closed-shell repulsive interaction on the overlap of the electron densities},
  author={Kim, Yung Sik and Kim, Seong Keun and Lee, Won Don},
  journal={Chemical Physics Letters},
  volume={80},
  number={3},
  pages={574--575},
  year={1981},
  publisher={Elsevier}
}

@article{bygrave_manby2012,
  title={The embedded many-body expansion for energetics of molecular crystals},
  author={Bygrave, PJ and Allan, NL and Manby, FR},
  journal={The Journal of Chemical Physics},
  volume={137},
  number={16},
  pages={164102},
  year={2012},
  publisher={AIP Publishing}
}

@article{ihm_1990_charge,
  title={Charge-overlap model of physical interactions and a combining rule for unlike systems},
  author={Ihm, G and Cole, MW and Toigo, Flavio and Klein, JR},
  journal={Physical Review A},
  volume={42},
  number={9},
  pages={5244},
  year={1990},
  publisher={APS}
}

@article{nobeli1998_different_K,
  title={Use of molecular overlap to predict intermolecular repulsion in N{\textperiodcentered}{\textperiodcentered}{\textperiodcentered} H—O hydrogen bonds},
  author={Nobeli, I and Price, SL and Wheatley, RJ},
  journal={Molecular Physics},
  volume={95},
  number={3},
  pages={525--537},
  year={1998},
  publisher={Taylor \& Francis}
}

@article{mitchell_price2000atomtypes,
  title={A systematic nonempirical method of deriving model intermolecular potentials for organic molecules: application to amides},
  author={Mitchell, John BO and Price, Sarah L},
  journal={The Journal of Physical Chemistry A},
  volume={104},
  number={46},
  pages={10958--10971},
  year={2000},
  publisher={ACS Publications}
}

@article{hirshfeld1977,
  title={Bonded-atom fragments for describing molecular charge densities},
  author={Hirshfeld, Fred L},
  journal={Theoretica Chimica Acta},
  volume={44},
  number={2},
  pages={129--138},
  year={1977},
  publisher={Springer}
}

@article{williams2001sapt_dft,
  title={Using Kohn- Sham orbitals in symmetry-adapted perturbation theory to investigate intermolecular interactions},
  author={Williams, Hayes L and Chabalowski, Cary F},
  journal={The Journal of Physical Chemistry A},
  volume={105},
  number={3},
  pages={646--659},
  year={2001},
  publisher={ACS Publications}
}

@article{misquitta2002sapt_dft,
  title={Intermolecular forces from asymptotically corrected density functional description of monomers},
  author={Misquitta, Alston J and Szalewicz, Krzysztof},
  journal={Chemical Physics Letters},
  volume={357},
  number={3-4},
  pages={301--306},
  year={2002},
  publisher={Elsevier}
}

@article{omol25,
  title={The open molecules 2025 (omol25) dataset, evaluations, and models},
  author={Levine, Daniel S and Shuaibi, Muhammed and Spotte-Smith, Evan Walter Clark and Taylor, Michael G and Hasyim, Muhammad R and Michel, Kyle and Batatia, Ilyes and Cs{\'a}nyi, G{\'a}bor and Dzamba, Misko and Eastman, Peter and others},
  journal={arXiv preprint arXiv:2505.08762},
  year={2025}
}

@article{spice,
  title={Spice, a dataset of drug-like molecules and peptides for training machine learning potentials},
  author={Eastman, Peter and Behara, Pavan Kumar and Dotson, David L and Galvelis, Raimondas and Herr, John E and Horton, Josh T and Mao, Yuezhi and Chodera, John D and Pritchard, Benjamin P and Wang, Yuanqing and others},
  journal={Scientific Data},
  volume={10},
  number={1},
  pages={11},
  year={2023},
  publisher={Nature Publishing Group UK London}
}

@article{zwanzig1988diffusion,
  title={Diffusion in a rough potential.},
  author={Zwanzig, Robert},
  journal={Proceedings of the National Academy of Sciences},
  volume={85},
  number={7},
  pages={2029--2030},
  year={1988}
}

@article{korona2010saptcc,
  title={Coupled cluster treatment of intramonomer correlation effects in intermolecular interactions},
  author={Korona, Tatiana},
  journal={Recent progress in coupled cluster methods: Theory and applications},
  pages={267--298},
  year={2010},
  publisher={Springer}
}

@article{patkowski2020review,
  title={Recent developments in symmetry-adapted perturbation theory},
  author={Patkowski, Konrad},
  journal={Wiley Interdisciplinary Reviews: Computational Molecular Science},
  volume={10},
  number={3},
  pages={e1452},
  year={2020},
  publisher={Wiley Online Library}
}

@article{cohen2012delocalization,
  title={Challenges for density functional theory},
  author={Cohen, Aron J and Mori-S{\'a}nchez, Paula and Yang, Weitao},
  journal={Chemical Reviews},
  volume={112},
  number={1},
  pages={289--320},
  year={2012},
  publisher={ACS Publications}
}

@article{johnson2023delocalization,
  title={Delocalization error: The greatest outstanding challenge in density-functional theory},
  author={Bryenton, Kyle R and Adeleke, Adebayo A and Dale, Stephen G and Johnson, Erin R},
  journal={Wiley Interdisciplinary Reviews: Computational Molecular Science},
  volume={13},
  number={2},
  pages={e1631},
  year={2023},
  publisher={Wiley Online Library}
}

@misc{SI,
  note = {See Supplementary Information at [URL will be inserted by publisher] }
}

@misc{so3lrv2,
  note = {This is an updated version of SO3LR that will be available at \href{https://github.com/general-molecular-simulations/so3lr} after further testing.}
}

@inproceedings{UL_HPC,
  title={Management of an academic HPC cluster: The UL experience},
  author={Varrette, S{\'e}bastien and Bouvry, Pascal and Cartiaux, Hyacinthe and Georgatos, Fotis},
  booktitle={2014 International Conference on High Performance Computing \& Simulation (HPCS)},
  pages={959--967},
  year={2014},
  organization={IEEE}
}

@article{feynman1939forces,
  title={Forces in molecules},
  author={Feynman, Richard Phillips},
  journal={Physical Review},
  volume={56},
  number={4},
  pages={340},
  year={1939},
  publisher={APS}
}

@article{2025qcml,
  title={The QCML dataset, Quantum chemistry reference data from 33.5 M DFT and 14.7 B semi-empirical calculations},
  author={Ganscha, Stefan and Unke, Oliver T and Ahlin, Daniel and Maennel, Hartmut and Kashubin, Sergii and M{\"u}ller, Klaus-Robert},
  journal={Scientific Data},
  volume={12},
  number={1},
  pages={406},
  year={2025},
  publisher={Nature Publishing Group UK London}
}




\end{document}


\maketitle
\section{Train/Test split of the DES15K dataset}
\begin{figure}[hbtp]
    \centering
        \includegraphics[width=3in]{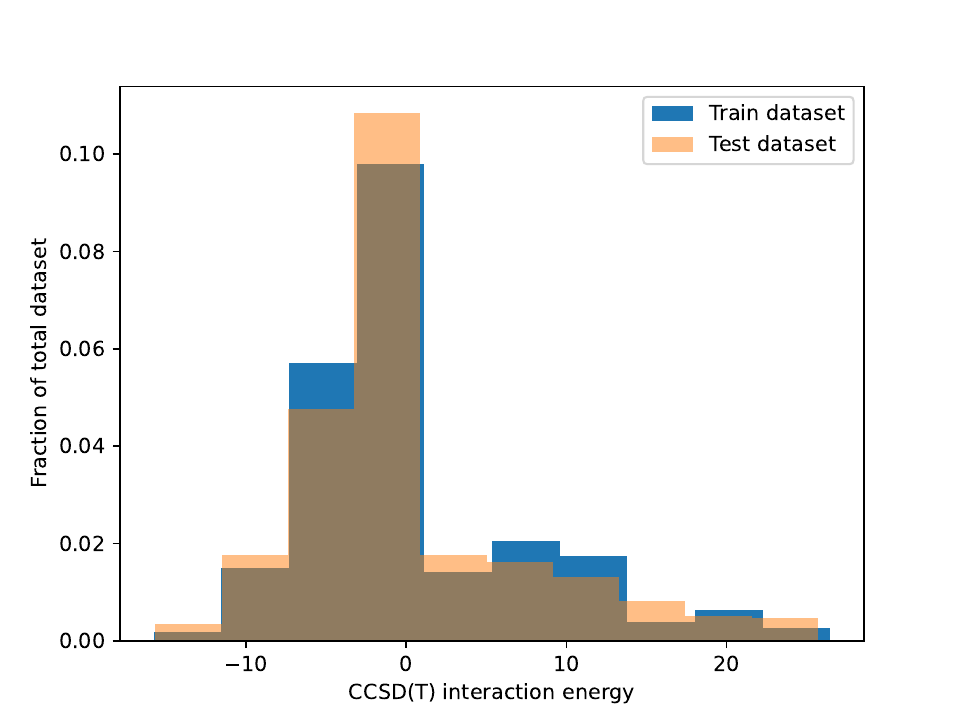}
        \hfill
    \caption{The distribution of interaction energies present in the training and test dataset splits of DES15K}
    \label{fig:train_test}
\end{figure}

We split the DES15K dataset \cite{des370k} into training and test datasets that are composed of distinct molecules in order to test transferablility of the DensIP model. Given that there are only 78 unique molecules in this set and the fact that small molecules tend to have smaller interaction energies, purely random training test splits often result in skewed sets that either grossly overestimate or underestimate the real error on the dataset. Therefore, we used 3-level stratification to create train/test splits that both represent the overall distribution of the dataset, see Fig. \ref{fig:train_test}.

\section{Learning curve and hyperparameter optimization}
Hyperparameter optimization and learning curves are produced by 5x5 crossfold validation. This means 5-fold cross validation is performed 5 separate times, in each instance the dataset is randomly split into folds. The 25 separate fits are then used to calculate the mean and standard deviation of the training and validation loss. The learning curve is produced by progressively including the dimers of each fold, which is kept fixed. Finally, all of the training data is used for the final fit and then evaluated on the test set as well as subsequent investigations detailed in the results section.

\begin{figure}[htbp]
    \centering
        \includegraphics[width=4in]{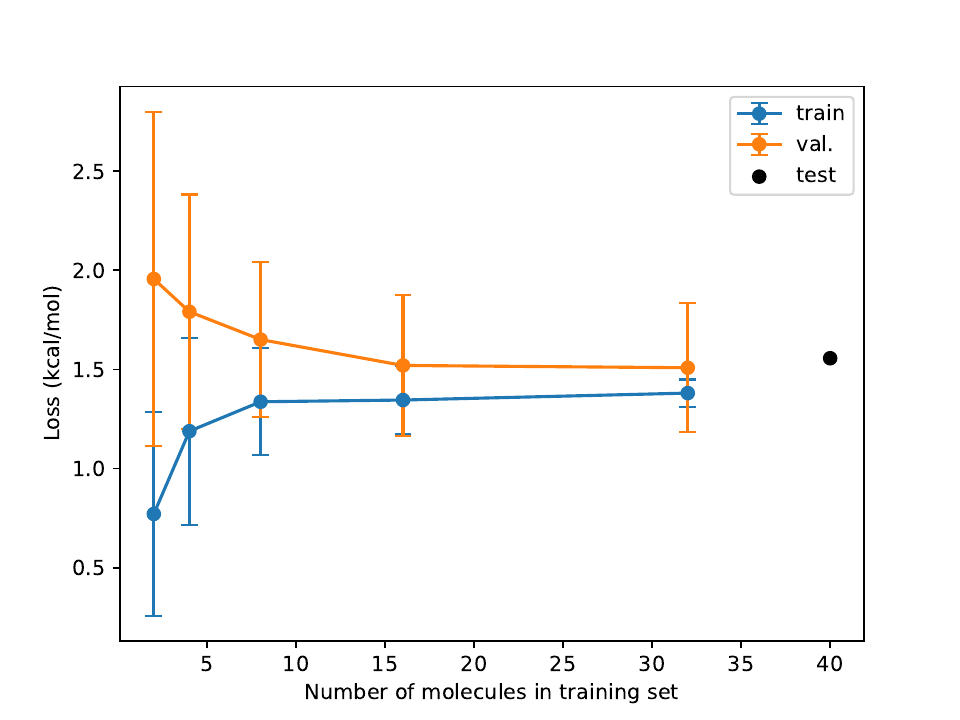}
        \hfill
    \caption{Learning curve for DensIP on the optimized configurations of DES15K}
    \label{fig:learning_curve}
\end{figure}

First, we present a learning curve for DensIP on the optimized dimer dataset, in Fig. \ref{fig:learning_curve}. Since the DensIP model consists of four parameters, the relatively small des15K dataset is nearly large enough to saturate the model. The validation sets only consist of 8 molecules and this is the source of the large uncertainty in the validation error.

\begin{figure}[htb]
    \centering
        \includegraphics[width=\textwidth]{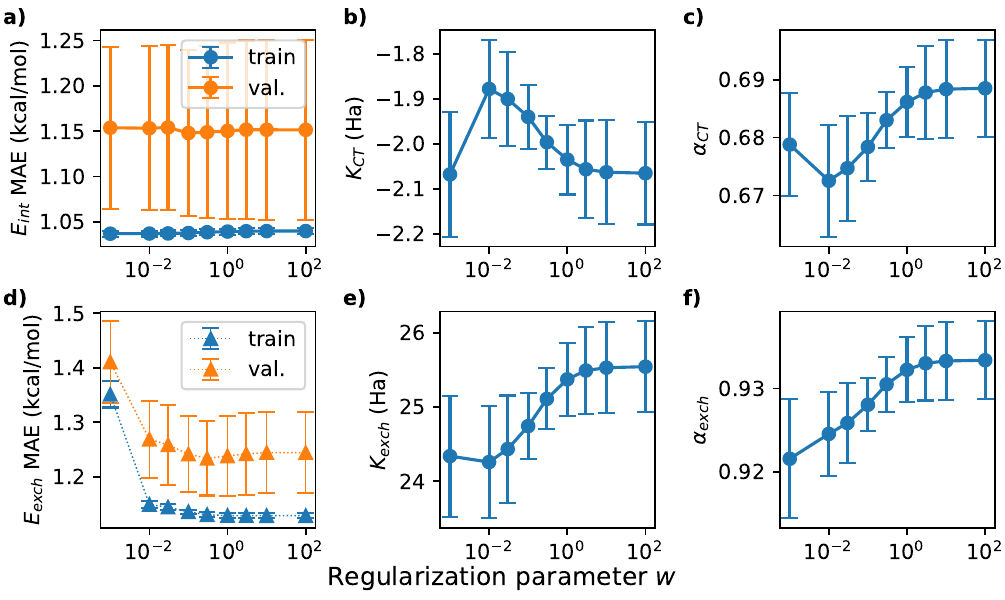}
        \hfill
    \caption{Optimizing the regularization parameter in the loss function to achieve physically-based model parameters}
    \label{fig:hyperparameters}
\end{figure}

Next, in Fig. \ref{fig:hyperparameters} we present a hyperparameter optimization of the $w$ parameter in the loss function (see eq. 8 of the main text). First, it can be seen in Fig. \ref{fig:hyperparameters}a that since the exchange and induction components are highly correlated, for any choice of exchange repulsion parameters, there exists a set of induction parameters such that the total energy of the model on the validation sets does not noticeably change. Given the non-uniqueness of energy decomposition methods, this does not impose an intrinsic problem, though in terms of interpretability it is more convenient to have the energy components that match a well defined method. It can be seen in Fig. \ref{fig:hyperparameters}d that increasing the regularization parameter causes the exchange-repulsion component to match SAPT0 exchange-repulsion and this also reduces the variability of the model parameters significantly for $w=1$. 

\section{DensIP with \textit{ab initio} densities}
We replace the ML densities predicted by DenSNet-ae and DenSNet-val with equivalent DFT densities using the PBE functional and an aug-cc-pVDZ basis set. The results are shown in Table \ref{tbl:DensIP_abiinitio}

\begin{table}[hb]
  \caption{RMSE of DensIP vs CCSD(T) on DES15K and PLF547 (kcal/mol) using either DenSNet predicted densities or DFT densities}
  \label{tbl:DensIP_abiinitio}
  \begin{tabular}{ccccccc}
    \hline
     &   Repulsive wall & Short-range & Equilibrium & Medium-range & MD & PLF-547\\
    \hline
   densip with $\rho_{ML}$ train      &  3.1 & 1.7 & 0.6 & 0.1 & -  & -  \\
densip with $\rho_{DFT}$ train    &  3.1 & 1.7 & 0.5 & 0.1  & - & -  \\

densip with $\rho_{ML}$ test  &   3.8 & 2.2 & 0.7 & 0.2 & 0.7 & 0.7 \\
densip with $\rho_{DFT}$ test   &  3.8 & 2.2 & 0.7 & 0.1 & 0.7  & 0.5  \\

    \hline
  \end{tabular}
  \end{table}


\section{Custom basis set}
The custom basis set is constructed from an uncontracted dense basis set developed in ref.~\cite{elking2010gmm}. Each element uses the same initial guess for gaussian widths, $\alpha$, (in Bohr$^{-2}$):
\clearpage
\begin{verbatim}
l=0
1305.9167743269209
365.7845449924546
126.0439336285438
49.43186555663881
21.21227962148198
9.721945897417026
4.678178313471295
2.3337760654836632
1.1966764301214545
0.627706688986485
0.33602608348653235
0.18312698182843895
0.10089277041403591
0.054770568510672205
0.029732707008
0.01614067344

l=1
126.0439336285438
49.43186555663881
21.21227962148198
9.721945897417026
4.678178313471295
2.3337760654836632
1.1966764301214545
0.627706688986485
0.33602608348653235
0.18312698182843895
0.10089277041403591
0.054770568510672205
0.029732707008
0.01614067344

l=2
4.678178313471295
2.3337760654836632
1.1966764301214545
0.627706688986485
0.33602608348653235
0.18312698182843895
0.10089277041403591
0.054770568510672205
0.029732707008
0.01614067344

l=3
4.678178313471295
2.3337760654836632
1.1966764301214545
0.627706688986485
0.33602608348653235
0.18312698182843895
0.10089277041403591
0.054770568510672205
0.029732707008
0.01614067344
\end{verbatim}
where the Gaussian is defined as $e^{-\alpha r^2}$. C, N, and O use the full definition, while H only includes orbitals up to l=2. As described in the main text, DenSNet learns optimal Gaussian widths and coefficients starting from these initial values. This was shown to yield overlaps with higher accuracy than using a basis set starting from jkfit basis set defaults for aug-cc-pVQZ \cite{weigend2002_jkfit}.

\section{DES15K composition}
For convenience, we also list the smiles strings by class of all molecules that appear in the subset of DES15K used in this study. The smiles strings that are in italics indicate molecules that are only in the optimized geometry dataset. All other strings are in both the optimized geometry dataset and the MD dataset.

\underline{alcohols}

CCCO, CCO, CC(O)C, CO, OC1CCCC1, OC1CCCCC1, OCCCCO, OCCCO, OCCO

\underline{aldehydes}

CC=O, C=O

\underline{alkanes}

C1CCCC1, C1CCCCC1, C, CC, CCC, CC(C)C, CCCC, CC(C)(C)C, CCCCC, CCCCCC

\underline{alkenes}

C=C, CC=C, CC=CC, CC(=C)C, CC=C(C)C, CC(=C(C)C)C

\underline{alkynes}

CC\#CC, CC\#C, C\#C, \textit{CCC\#C}

\underline{amides}

CC(=O)N, CC(=O)N(C)C, CNC=O, CNC(=O)C, NC=O, O=CN(C)C

\underline{amines}

C1CCCN1, C1CCCNC1, CCN, CCN(C)C, CN, CNC, CN(C)C, CNCC, N

\underline{arenes}

c1ccccc1, Cc1ccccc1, Oc1ccccc1

\underline{carboxylic acids}

CC(=O)O, OC=O

\underline{esters}

COC(=O)C, COC=O

\underline{ethers}

C1CCCO1, C1CCCOC1, C1CCOCO1, C1OCCO1, CCCOC, CCOCC, COCC, COC, COCOC, O1CCOCC1, O1COCOC1

\underline{H$_2$}

\textit{[H][H]}

\underline{ketones}

CC(=O)C

\underline{N-heteroarenes}

c1cccnc1, c1ccncn1, n1ccncc1, c1ccc2c(c1)[nH]cc2, c1ccc[nH]1, c1ncc[nH]1, Cc1cnc[nH]1, Cc1c[nH]cn1

\underline{nitriles}

CC\#N, C\#N, \textit{CCC\#N}

\underline{water}

O
\printbibliography